\documentclass[12pt,preprint]{aastex}

\usepackage{graphicx}
\usepackage{amsmath}
\usepackage{color}
\usepackage{xspace}

\newcommand{\pan}{{\sc Pandurata}\xspace}
\newcommand{\harm}{{\sc Harm3d}\xspace}

\shorttitle{Disk Emission from Spinning Black Holes}
\shortauthors{Schnittman, Krolik, \& Noble}

\begin{document}

\title{Disk Emission from Magneto-hydrodynamic Simulations of Spinning
  Black Holes} 

\author{Jeremy D.\ Schnittman}
\affil{NASA Goddard Space Flight Center \\
Greenbelt, MD 20771}
\email{jeremy.schnittman@nasa.gov}
\and
\author{Julian H.\ Krolik}
\affil{Department of Physics and Astronomy, \\
Johns Hopkins University\\
Baltimore, MD 21218}
\email{jhk@pha.jhu.edu}
\and
\author{Scott C.\ Noble}
\affil{Department of Physics and Engineering Physics, \\
University of Tulsa\\
Tulsa, OK 74104}
\email{scott-noble@utulsa.edu}

\begin{abstract}
We present the results of a new series of global 3D relativistic
magneto-hydrodynamic (MHD) simulations of thin accretion disks around
spinning black holes. The disks have aspect ratios of $H/R\sim 0.05$
and spin parameters $a/M=0, 0.5, 0.9$, and $0.99$. Using the
ray-tracing code \pan, we generate broad-band thermal spectra and
polarization signatures from the MHD simulations. We find that the
simulated spectra
can be well fit with a simple, universal emissivity profile that
better reproduces the behavior of the emission from the inner disk,
compared to traditional analyses carried out using a Novikov-Thorne
thin disk model. Lastly, we show how spectropolarization observations
can be used to convincingly break the spin-inclination degeneracy
well-known to the continuum fitting method of measuring black hole
spin. 
\end{abstract}


\section{Introduction}\label{section:intro}

Recent years have seen great strides in our ability to simulate global
properties of accretion disks in a fashion dependent on actual
physical mechanisms rather than {\it ad hoc} phenomenological
models. Beginning a decade ago, several algorithms have been developed
capable of treating 3D MHD in full general relativity
\citep{DeVH2003,Gammie2003,Noble2009,Penna2010}. With those algorithms, it has
become possible to study the bolometric energy budget of radiatively
efficient accretion
onto black holes \citep{Noble2009,Penna2010,Avara2015}, and from it predict
details of the spectra of stellar-mass black holes both in the
thermal-dominant state \citep{noble:11,kulkarni:11,Zhu2012} and in other
spectral states \citep{schnittman:13a}.
Potentially, fits of these predictions to measured spectra taken in
the thermal state could be used to constrain the spin of the black
hole and the inclination of the disk
\citep{Davis2011,Gou2011,Steiner2011,Gou2014}. 

However, to make these measurements quantitative requires careful
attention to a variety of sources of systematic error. Possible
contributors to the error budget include uncertainties about
atmospheric effects, both LTE and non-LTE, the disk inclination angle
(spectral fitting leaves spin and inclination degenerate), the
vertical structure of disks, and the intrinsic surface brightness
profile of the disk. The two latter issues are both amenable to study
through simulations, but both also exhibit subtle sensitivities to
algorithm (e.g., the consequences of internal radiation pressure:
\cite{Hirose2009,Jiang2013}) and grid resolution. In particular,
after the first generation of global energy budget simulations 
were completed, \citet{Hawley2011} showed that only one of them
(ThinHR: \cite{Noble2010}) even came close to having the spatial
resolution required for a proper description of nonlinear MHD
turbulence in the disk context; the others all fell far short.
Unfortunately, the Noble et~al. work treated only a Schwarzschild
spacetime and therefore gave little aid to constraining spin.

It is the first goal of this paper to present three new simulations in
Kerr spacetime (spin parameters $a/M = 0.5$, 0.9, and 0.99) with
resolution and vertical scale height comparable to that of the Noble et~al. Schwarzschild 
simulation. With the resulting data, we have constructed predictions of
the surface brightness profile and the resulting thermal spectra and
compare them to previous results. Additionally, we make predictions of
spin-dependent effects in the polarization spectrum of the thermal
state, and show how spectropolarization observations can break
degeneracies that have traditionally plagued the continuum fitting
method for measuring spin. Lastly, we present a new analytic model for
the emissivity profile of disks in the thermal-dominant state,
applicable to {\it any} value of $a/M$. 

In general, we find these new simulations lead to results similar to
the non-spinning case described in
\citet{noble:11}. Namely, the MHD disks produce a significant amount
of additional flux in the region around the inner-most
stable circular orbit (ISCO). This additional luminosity is manifest
in a small, but consistent, increase in flux at the high-energy tail
of the thermal spectrum as measured by a distant observer. Fitting the
simulated spectra to traditional analytic thin disk models (e.g.,
\citet{novikov:73}) therefore leads to small but significant bias
towards higher spin parameters. Our proposed emissivity profile
provides a better overall fit to the data, as well as generally a more
faithful inference of the spin parameter.

\section{Simulation Data}\label{section:harm}

\subsection{Global fluid properties}\label{section:global}

Our simulations were carried out with the \harm code \citep{Noble2009}
and were designed to resemble closely the ThinHR Schwarzschild
simulation of \cite{Noble2010}. All used the same cooling function and
target temperature profile, designed to keep the disk aspect ratio
$H/r \simeq 0.05$. All also began with the same initial condition, a
hydrostatic torus with a pressure maximum at $r=35M$ and an inner edge
at $r=20M$ (here and subsequently, we use units in which $G=c=1$).
Threading these tori there was a dipolar magnetic field with mean
initial plasma $\beta = 100$. The spatial grids for the three Kerr
spacetime simulations were very similar to the one utilized for the
Schwarzschild case; the only alteration was to extend the radial grid
to smaller Kerr-Schild radii in order to follow the shrinkage of the
event horizon with increasing spin. As in the earlier work, our
radial grid was logarithmic (i.e., constant $\Delta r/r \simeq 0.004$) 
and stretched from just inside (5 cells) the event horizon to $r =
70M$. Consequently, there were 912, 960, 1020, and 1056 radial cells
in the simulations with $a/M = 0$, 0.5, 0.9, and 0.99,
respectively. In all four simulations, there were 160 cells in the
polar angle direction, and these cells were strongly concentrated to
the equatorial plane. Likewise in all cases, the simulation domain
included only the 1/4 circle from $\phi = 0$ to $\phi = \pi/2$, with
64 cells across that range and periodic boundary conditions that
mapped cells at $\phi = 0$ to $\phi = \pi/2$. Boundary conditions on
the inner and outer radial surfaces were outflow; the polar-angle
boundaries at the axis ($\theta = 0,\pi$) were continuous across the
pole. 

The ThinHR simulation was run to a time 15,000$M$; the Kerr simulations
were run to $\simeq$17,000 -- 18,000$M$. As is generally the case, inflow
equilibrium, to the degree it is achieved, begins at the ISCO and
gradually moves outward. It can be characterized in (at least) two
ways: by a statistical steady-state in the accretion rate onto the
black hole and by constancy in the accretion rate as a function of
radius when time-averaged over the period of interest. Here we show
both measures. 

\begin{figure}
\caption{\label{fig:mdot_t} Accretion rate as a function of time for
  all four simulations; from left to right and then top to bottom,
  they are $a/M = 0.0$, 0.5, 0.9, and 0.99. The units of accretion are
  fractions of the initial mass on the grid per $M$.} 
\begin{center}
\includegraphics[width=0.45\textwidth]{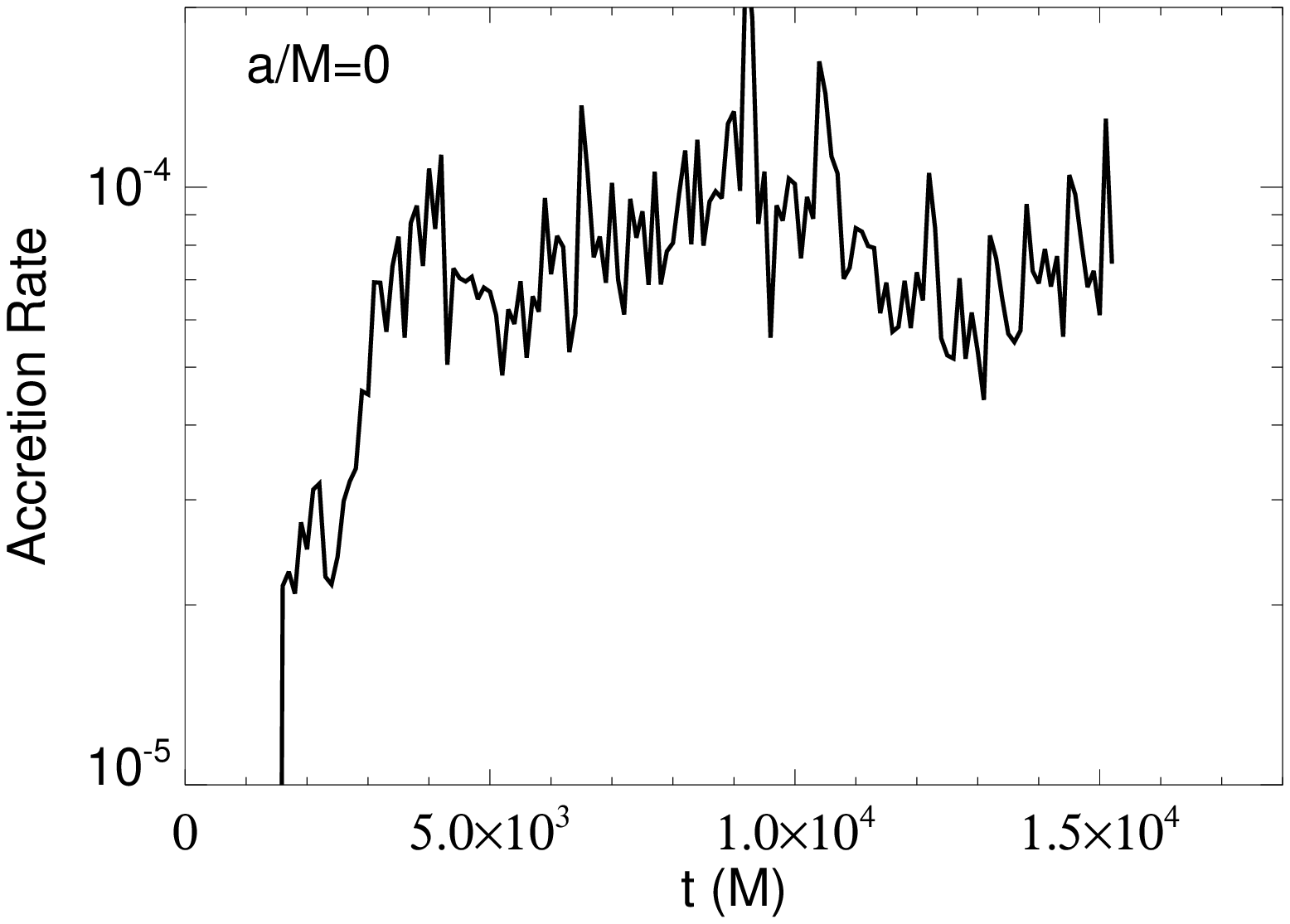}
\includegraphics[width=0.45\textwidth]{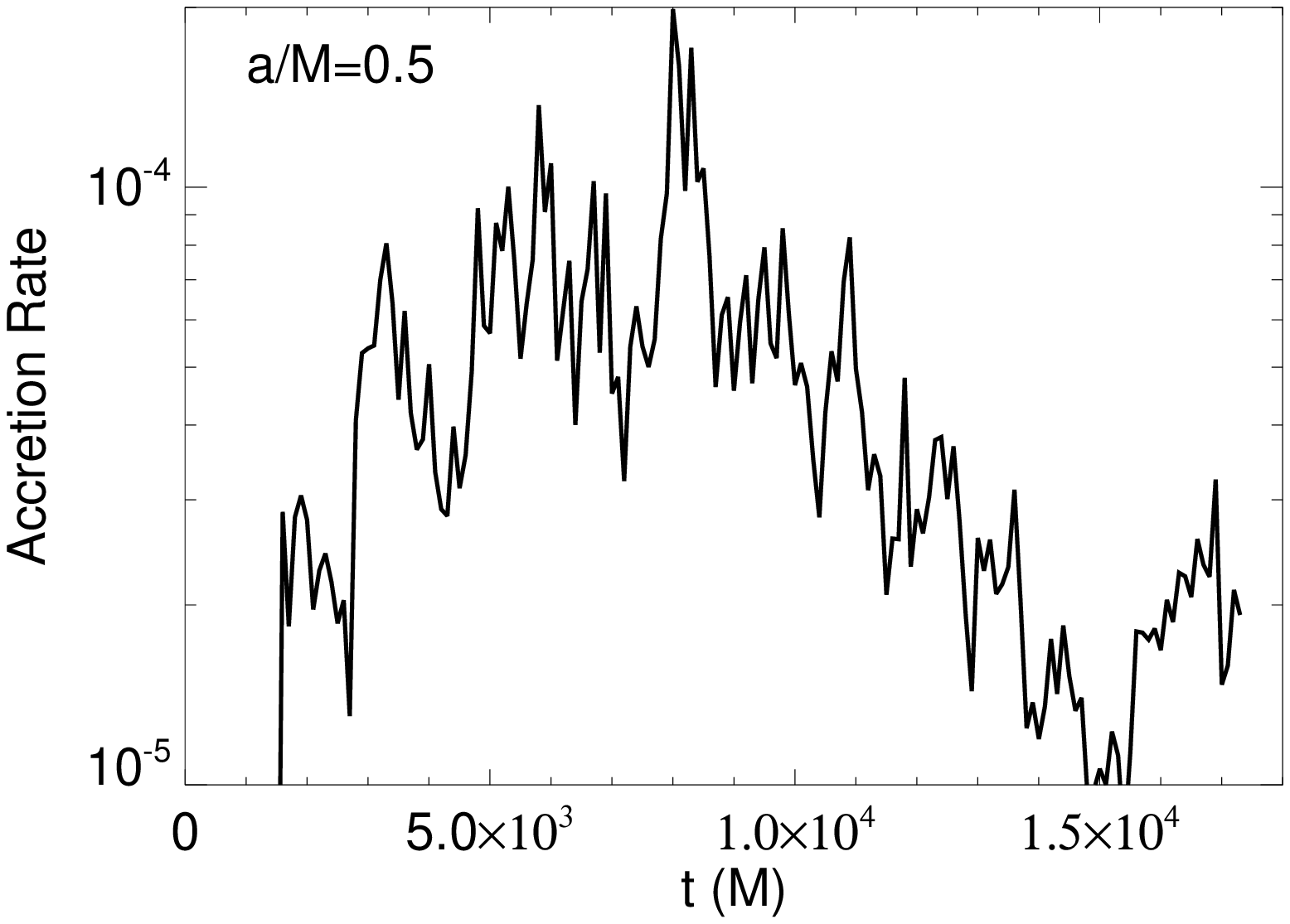}\\
\includegraphics[width=0.45\textwidth]{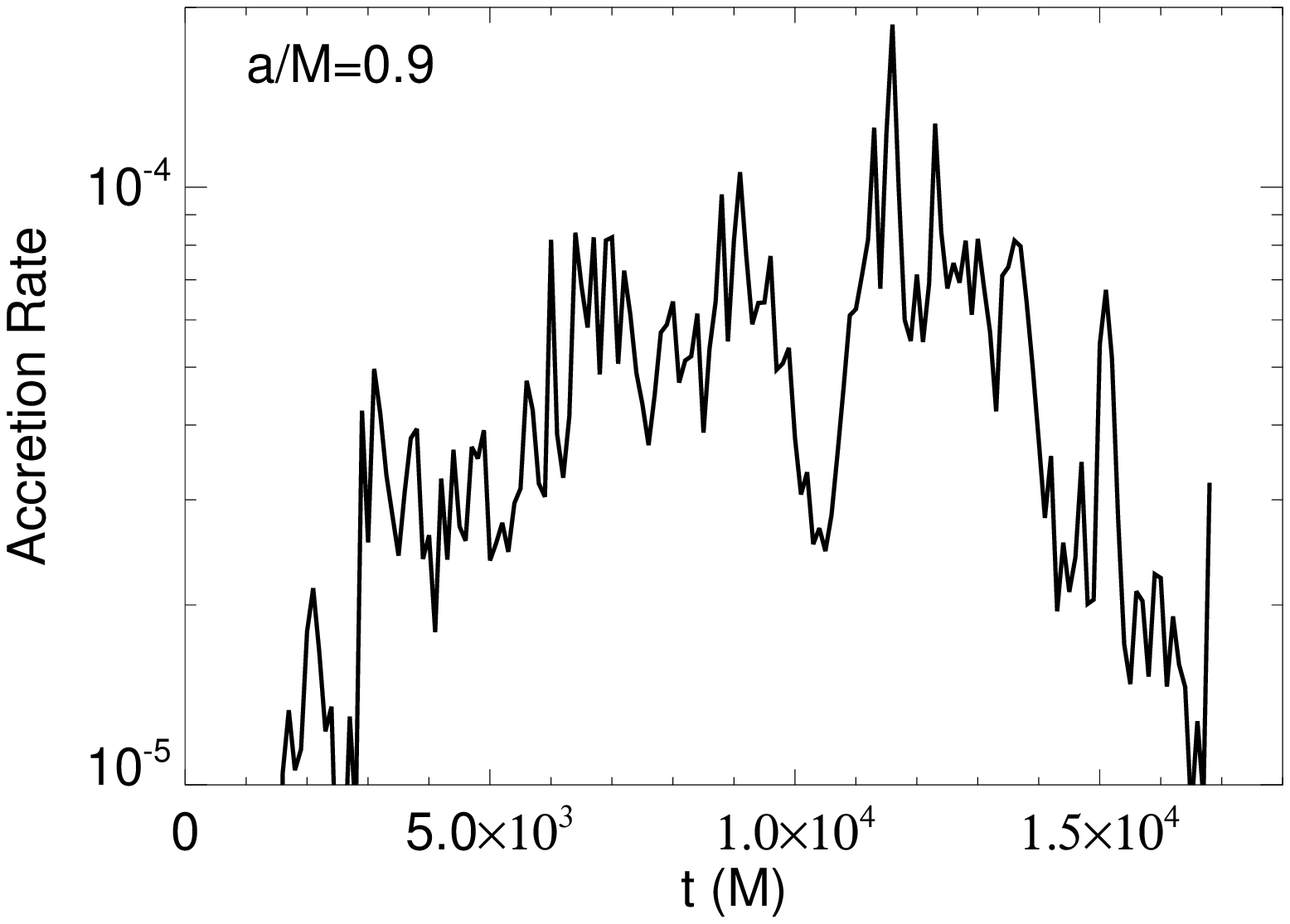}
\includegraphics[width=0.45\textwidth]{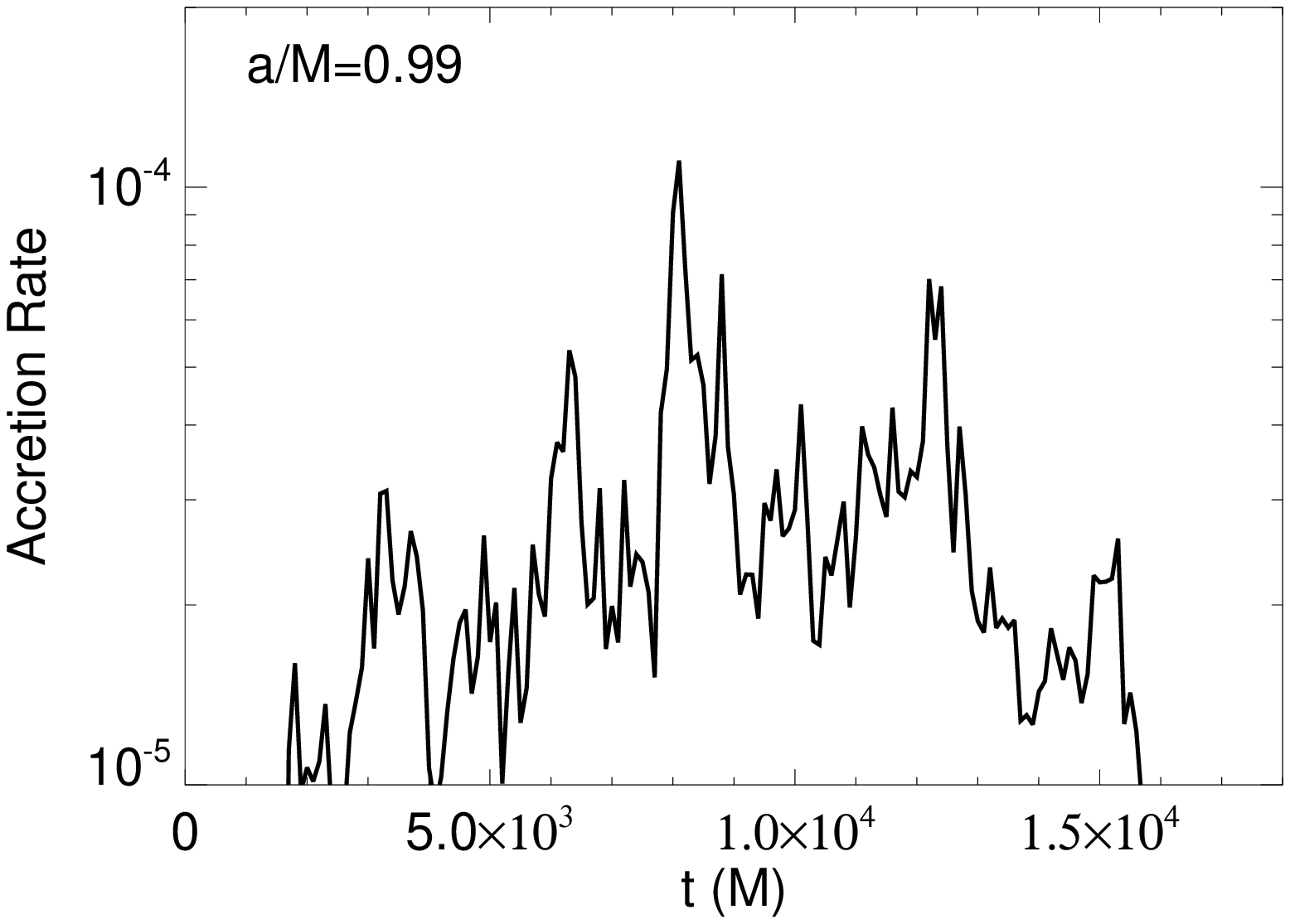}
\end{center}
\end{figure}

Figure~\ref{fig:mdot_t} shows how the accretion rate through the event
horizon varied over time in all four simulations. The peak accretion
rates were similar in all four simulations, but, particularly at late
times, there is a tendency for the accretion rate to diminish with
increasing spin. This tendency was also seen in earlier global
simulations with thicker disks and shorter durations
\citep{krolik:05}. Because the period before 10,000$M$ tends to be
dominated by transients associated with initial channel solutions, we
have restricted our analysis of disk output to periods after that
time, 10,000 -- 15,000$M$ for $a/M = 0$, 0.5, and 0.9, 12,000 --
15,000$M$ for $a/M = 0.99$.

\begin{figure}
\caption{\label{fig:mdot_r} Time-averaged accretion rate as a function
  of radius in ratio to the rate at the horizon for all four
  simulations; from left to right and then top to bottom, they are
  $a/M = 0.0$, 0.5, 0.9, and 0.99. The unit of radius in each case is
  the radial coordinate of the ISCO for that spin.} 
\begin{center}
\includegraphics[width=0.45\textwidth]{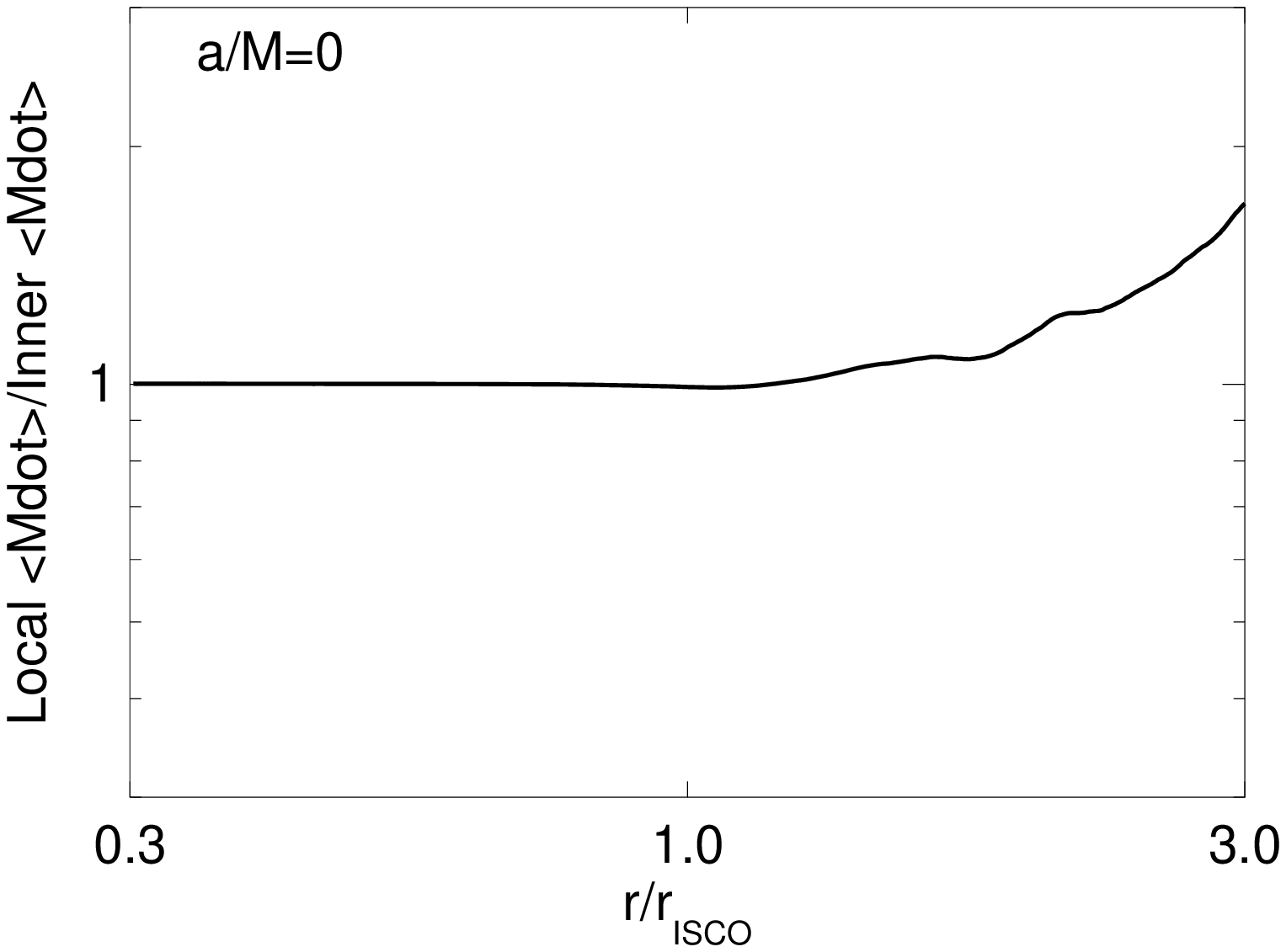}
\includegraphics[width=0.45\textwidth]{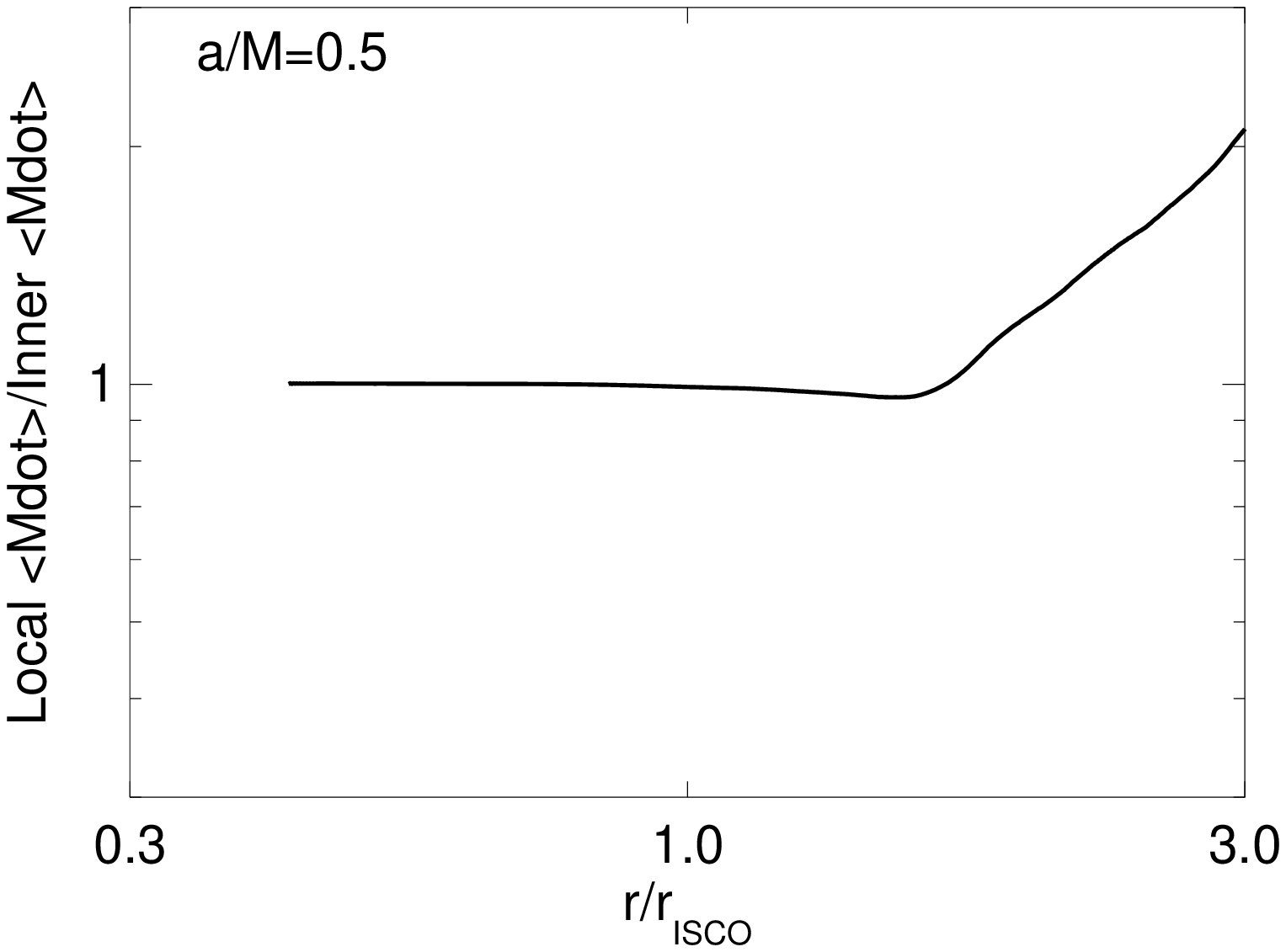}\\
\includegraphics[width=0.45\textwidth]{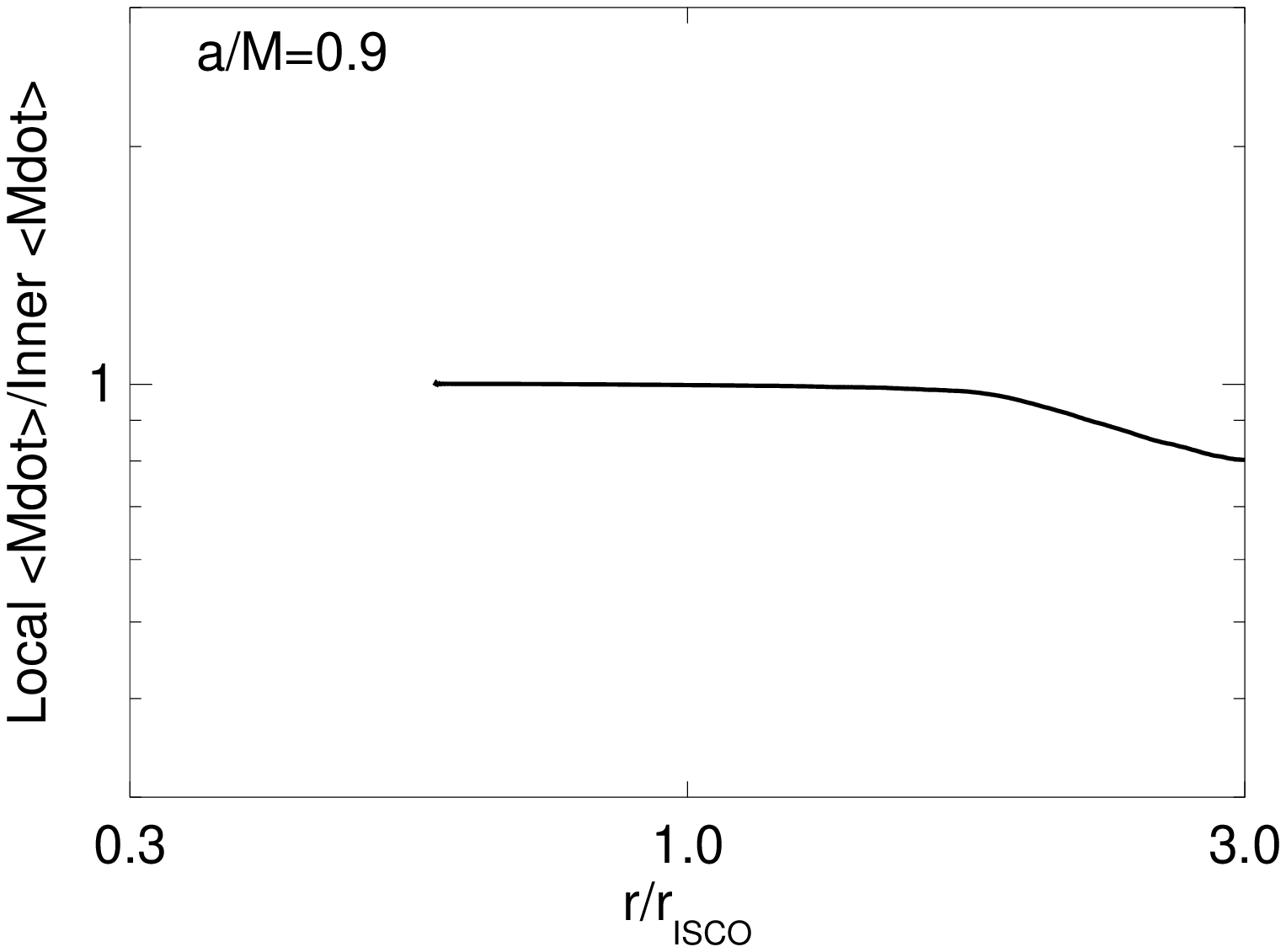}
\includegraphics[width=0.45\textwidth]{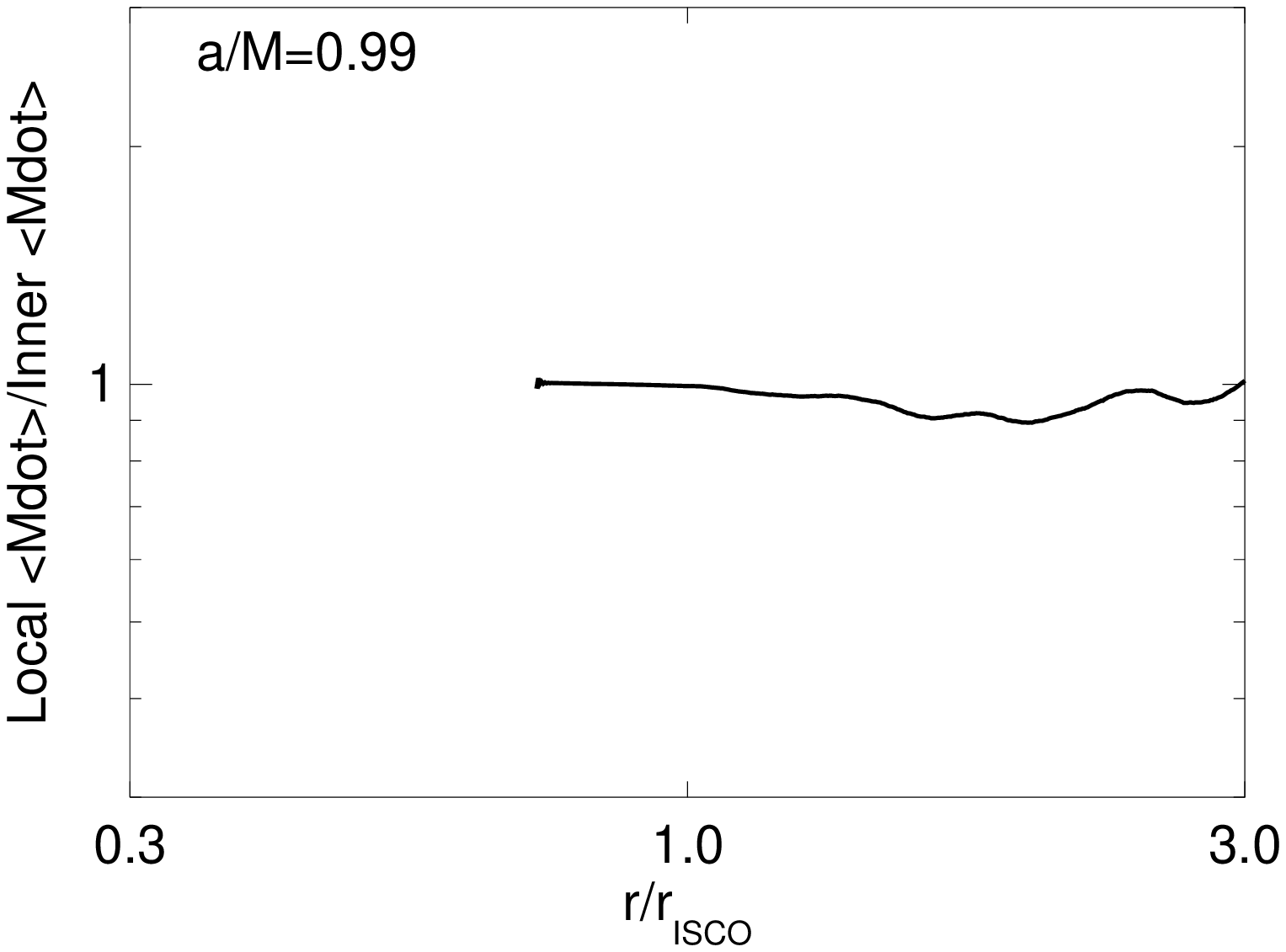}
\end{center}
\end{figure}

Despite large variations in the time-dependent accretion rate at
the horizon, the time-averaged accretion rate had relatively little
variation in radius, a sign that overall there was little change in
the disk's mass distribution. To illustrate that fact, in
Figure~\ref{fig:mdot_r} we show the ratio $\langle \dot M
(r)\rangle/\langle \dot M(r_H)\rangle$ as a function of $r/r_{\rm
  ISCO}$. Both axes are logarithmic in order to emphasize that ratios
are what matter most in this context. Inflow equilibrium is best
judged in terms of fractional departures in the accretion rate; radial
variation is best considered in units of $r_{\rm ISCO}$ because the
greatest energy release takes place at or near the ISCO and, in the
Newtonian limit, the potential is a power-law in radius. In
three of the four cases ($a/M = 0$, 0.9, and 0.99), the time-averaged
accretion rate changes by at most a few tens of percent for $r \lesssim 3 r_{\rm ISCO}$.
In the case of $a/M = 0.5$, this criterion is achieved only out to $r
\simeq 2 r_{\rm ISCO}$, beyond which it rises to about twice the
accretion rate at the horizon.

Over these time periods, the MRI quality factors for all four
simulations were quite good. We measured the quality factors $Q_z$
(the mean number of cells per fastest growing vertical wavelength) and
$Q_\phi$ (the mean number of cells per fastest growing toroidal
azimuthal wavelength) by constructing mass-weighted averages on
spherical shells and then time-averaging over the same periods for
which inflow equilibrium was evaluated \citep{Noble2010,Hawley2011}. When $a/M=0$, $Q_z \simeq
25$ and $15 \lesssim Q_\phi \lesssim 20$ across the entire range of
radii; when $a/M = 0.5$, $15 \lesssim Q_z \lesssim 30$ and $10
\lesssim Q_\phi \lesssim 25$, with comparable quality factors for
$a/M=0.9$ and $a/M=0.99$.  These numbers compare very favorably
with the standard set by \cite{Hawley2011} and \cite{Sorathia2012}
that for proper description of the nonlinear development of MHD
turbulence in accretion disks if $Q_\phi \gtrsim 20$, then $Q_z$
should be $\gtrsim 10$. 

\subsection{Radiation}\label{section:radiation}

Given the density, cooling rate, and four-velocity of the fluid at
each point in the simulation, we construct a disk surface temperature map as in
\citet{noble:11}.  For the purpose of ray-tracing to infinity, we remove the
geometric effects of finite disk thickness, assuming that all the thermal flux
comes from a thin disk near the midplane, while still preserving the time and
azimuthal variability in the simulation data. \citet{kulkarni:11} do
the former, but not the latter, leading to systematically softer
spectra. The radiation is emitted from this surface with
the limb-darkening and polarization properties appropriate for a
scattering-dominated, optically thick atmosphere \citep{chandra:60}.

Since we are in this paper primarily focused on the thermal state, in
which the coronal Compton-$y$ parameter is expected to be small,
we ignore any scattering in the corona (to a large degree, in the
thermal state, the corona
can simply be thought of as an extension of the accretion disk
atmosphere). The photons therefore follow geodesic paths
from the disk until they reach the distant observer or are captured by
the event horizon. However, because we are also
interested in polarization signatures, we include scattering of
returning radiation off the optically-thick disk body, as in
\citet{schnittman:09}. 

As a test of this method, we have also tried using emission models
that include only the disk proper, as defined by the region between
electron-scattering photospheres, as in \citet{schnittman:13a}. This
approach requires specifying a physical accretion rate in order to fix
the density scale in cgs units. For $\dot{m}=0.03$ in Eddington units
(roughly the lower limit of the thermal-dominant state), we find that,
for the $a/M=0$ run, roughly $80\%$ of the dissipation takes place in
the disk proper, with that number rising to $90\%$ for the spinning
cases. For all cases, we find that the coronal and disk radial
emissivity profiles are nearly identical. Thus, if we were to
include only the disk emission, the thermal luminosity would
decrease slightly, but the spectral shape remains unchanged.

\section{Results}\label{section:results}

\subsection{Efficiency}\label{sec:efficiency}

One of the most basic results that comes from these global MHD
simulations is the total radiative efficiency of the thin accretion
disk. Astrophysically, this number is critical for connecting the
accretion history and growth of super-massive black holes in the
universe with the observed luminosity and mass distribution functions
\citep{soltan:82}. While it is almost impossible to measure the
efficiency of an individual black hole, the shape of the thermal
spectrum (which of course is observable) is empirically related to the
radiative efficiency because higher efficiency is typically due to
higher emission at small radii, producing higher temperatures and
harder spectra. 

In Table \ref{table:eta} we show several measures of the radiative
efficiency. We define $\eta$ (radiated) to be the ratio between the
time-averaged luminosity emitted by the disk
and the time-averaged rest-mass accretion rate (photon energy and
observer time both measured at infinity). Similarly, $\eta$ (captured)
is the ratio of the time-averaged luminosity crossing the horizon
to the time-averaged accretion rate (for photons crossing the horizon,
the energy-at-infinity is easy to measure: $E_\infty=-p_t$). Thus the
net radiative efficiency measured at infinity is
$\eta$(radiated)-$\eta$(captured). 

$\eta_{\rm NT}$ is the classical Novikov-Thorne efficiency, and
$\eta_{\rm Harm3d}$ is calculated from our simulations. 
Note that $\eta_{\rm NT}$ (radiated) is identical to the
specific binding energy of a test particle orbiting at the ISCO.
Lastly, for the simulation data, we also show $\eta_{\rm Harm3d}$
(advected), a measure of the accreted energy carried into
the hole in the form of heat and magnetic field. In principle, this
energy could be liberated via additional radiation mechanisms. 

As first shown in \citet{thorne:74}, we find that very little radiation is
actually captured by the black hole. Although the additional flux
emitted near and inside the ISCO leads to somewhat higher capture
fractions than the NT model would predict, it is still less than $5\%$ of the
total emitted flux even for $a/M = 0.99$. In the end, the net radiative
efficiency is quite close to $\eta_{\rm NT}$, as the additional MHD
dissipation is negated by capture and advection, consistent with
earlier work \citep{noble:11,kulkarni:11}. 
As shown in \citet{noble:11}, much of the increase in
radiative efficiency relative to NT is due to the enhanced emission
near the ISCO, both immediately inside and outside that radius.

In a certain sense, our figures for the efficiency are conservative
estimates. By adopting an initial condition with nested dipolar field
loops, we assure that there is some net magnetic flux trapped on the
event horizon, but in terms of the $\Upsilon$ measure introduced by
\cite{Gammie1999}, it is smaller than the maximum that can be
achieved. Here we use the definition of \cite{McKinney2012}, in which
$\Upsilon \equiv 0.2 \Phi/(r_g\sqrt{\dot Mc})$, for accretion rate
$\dot{M}$, gravitational radius $r_g$, and $\Phi$ the integral of the
absolute value of the radial magnetic field around a spherical surface
containing the black hole. 

When $\Upsilon \simeq 6$ (roughly $3\times$ the value in our
simulation), \cite{Avara2015} found that the radiative efficiency of a
disk with somewhat greater aspect ratio ($h/r \simeq 0.1$) orbiting in
an $a/M=0.5$ spacetime was roughly double what we found for the same
spin parameter.   This contrast suggests that the magnetic
augmentation to radiative efficiency rises very sharply over 
the range in $\Upsilon$ between these values. However, in thermal
states of black hole binaries, which do not show evidence of radio
jets, one might not expect such large values of $\Upsilon$. Moreover,
the intrinsic field topology in accreting black holes is entirely
unknown. Other sorts of field topology---higher-order multipoles in
the poloidal field or intrinsically toroidal field---may exist, and
how the efficiency behaves as a function of net toroidal flux, etc.,
remains to be explored. All that can be said at the moment is that for
$\Upsilon$ comparable to or smaller than in our simulations, the
internal magnetic stresses depend only weakly on the field topology
\citep{beckwith:08}. 

We see significant variability of the accretion rate
and radiative efficiency within each simulation, as well as
qualitative differences from one simulation
to the next (e.g., differences in the radial fluid properties that do
not appear to vary monotonically with spin), representative of the
stochastic nature of the turbulent 
MHD disks. In principle, we could carry out an ensemble of runs for
each spin to measure both the mean radiative efficiency and the
variance of the observed efficiency due to the random fluctuations in
the simulations. However, even the four simulations we did carry out
were extremely expensive computationally, and any more would be very
difficult to justify.

\begin{table}[ht]
\caption{\label{table:eta} For a range of spins, the emitted
  bolometric radiative efficiency $\eta$ (radiated) and the fraction of
  flux captured by the black hole $\eta$ (captured), for both NT disks
  and \harm simulations. Also listed is the fraction of advected
  thermal and magnetic energy in the MHD simulations.}
\begin{center}
\begin{tabular}{lccccc}
\hline
\hline
$a/M$ & $\eta_{\rm NT}$ & $\eta_{\rm NT}$ & $\eta_{\rm Harm3d}$ &
$\eta_{\rm Harm3d}$ & $\eta_{\rm Harm3d}$ \\ 
 & (radiated) & (captured) & (radiated) & (captured) & (advected) \\ 
\hline
 0.0  & 0.0572 & 0.0001 & 0.0607 & 0.0020 & 0.0080 \\
 0.5  & 0.0821 & 0.0005 & 0.0833 & 0.0016 & 0.0120 \\
0.9  & 0.1558 & 0.0027 & 0.1600 & 0.0048 & 0.0115 \\
0.99 & 0.2640 & 0.0087 & 0.2461 & 0.0111 & 0.0220 \\
\end{tabular}
\end{center}
\end{table}

\subsection{Radial luminosity profile}\label{sec:profile}

\begin{figure}
\caption{\label{fig:harm_nt} Luminosity profile $dL/dr$, integrated
  over $\theta$ and $\phi$, and averaging over time.
  The solid red curve is the \harm emission, the
  dashed curve is the flux reaching infinity, and the dotted curve is
  the flux captured by the black hole. The solid black curve corresponds
  to the Novikov-Thorne prediction for emission. For all cases, the
  Eddington-normalized accretion rate is $\dot{m}=0.1$.}
\begin{center}
\includegraphics[width=0.45\textwidth]{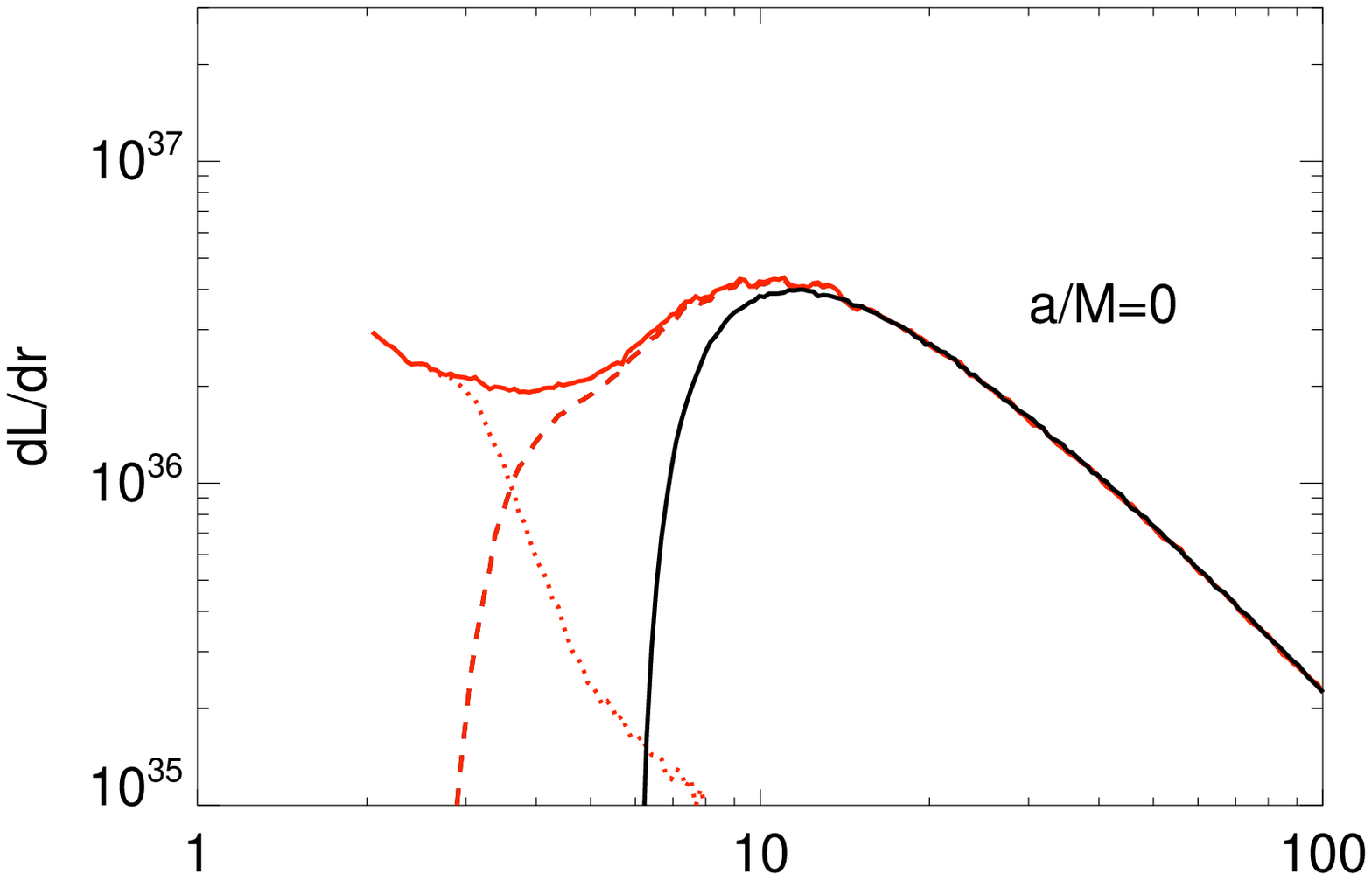}
\includegraphics[width=0.45\textwidth]{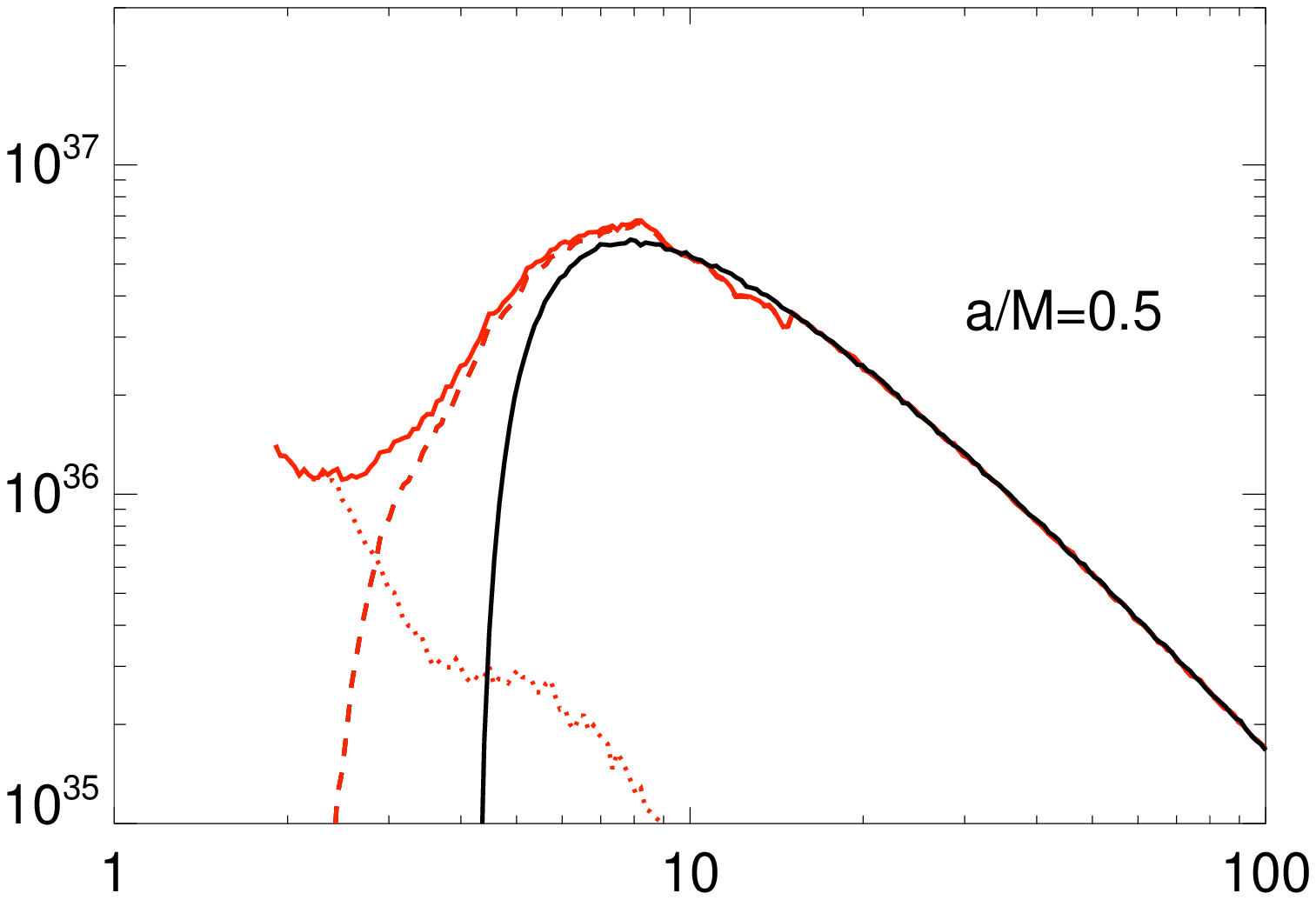}\\
\includegraphics[width=0.45\textwidth]{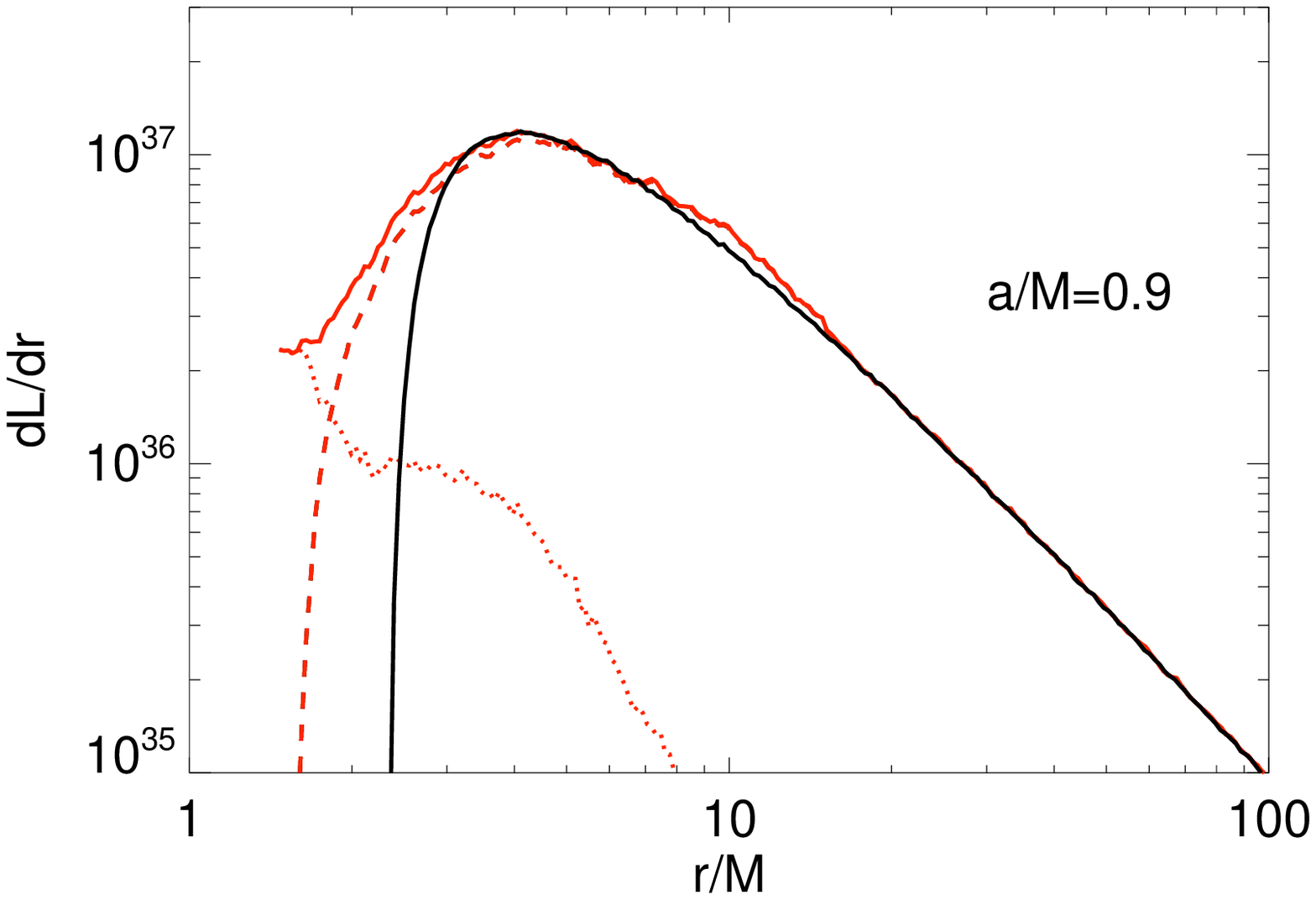}
\includegraphics[width=0.45\textwidth]{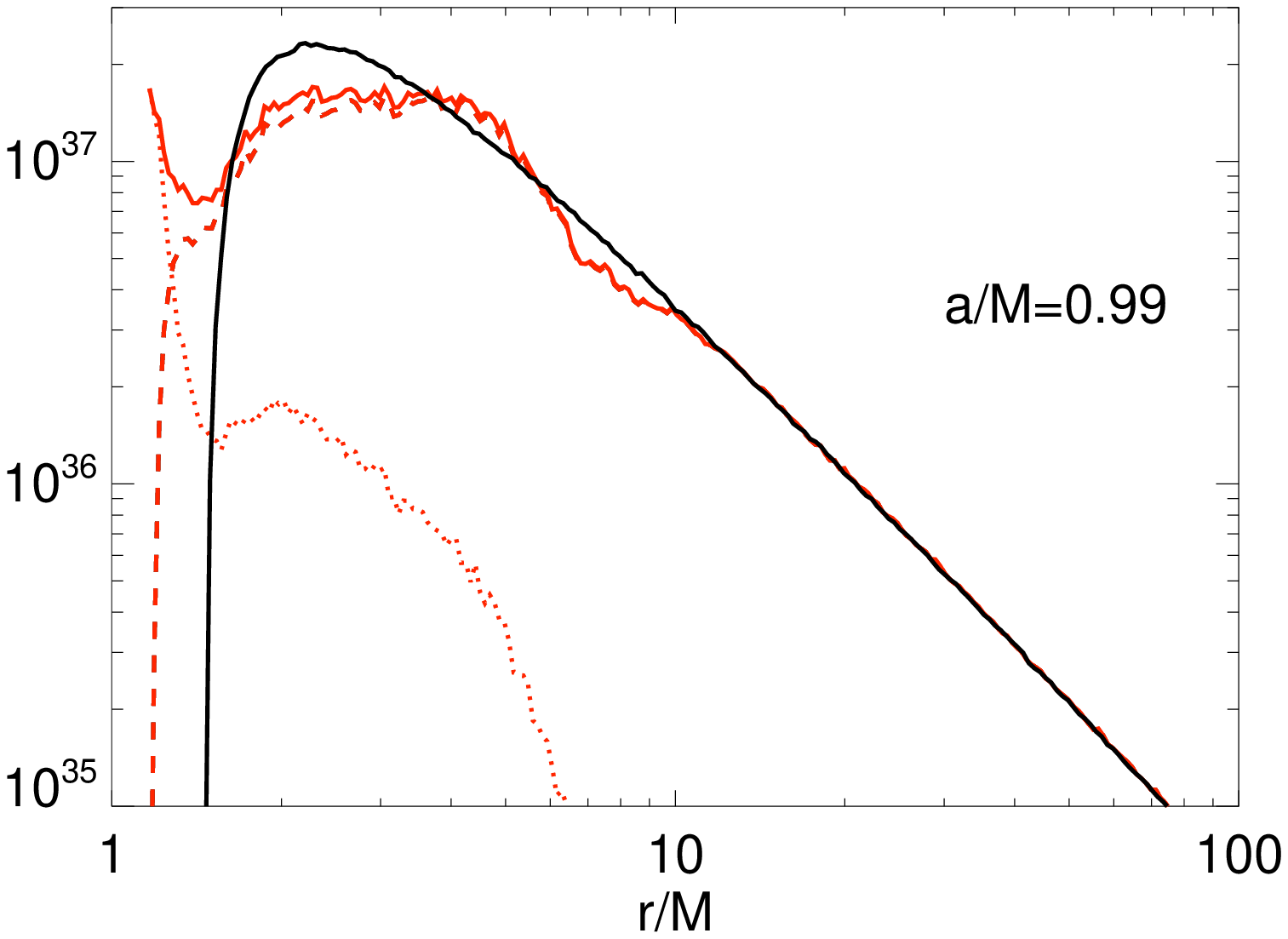}
\end{center}
\end{figure}

In Figure \ref{fig:harm_nt} we plot the radial luminosity profile
$dL/dr$ for each of the black hole spins. In each plot, the solid
red curve represents the time- and angle-averaged emitted flux from
the \harm simulation data. The red dashed curve is the flux that
reaches infinity, and the red dotted curve is the flux
captured by the black hole. The solid black curve is the emitted flux
from an NT disk.

All four cases share a number of qualitative features.   The emission per unit
radial coordinate peaks a short distance outside the ISCO, then
decreases slightly before leveling off in the plunging region. Inside a
point roughly halfway between the ISCO and the horizon, the flux emitted by
the inward-flowing gas is almost entirely captured by the black
hole.   Contrasts also exist.   For example, compared to the others, the fraction
of the luminosity emitted in the $a/M=0.5$ plunging region is relatively small.
The fluctuations alluded to earlier are large enough that these contrasts may
be due to our comparatively short averaging times.

Because the ISCO position in Boyer-Lindquist coordinates is a function of spin,
detailed comparison of the four cases' radial profiles requires defining a
new radial coordinate relative to which the emission profiles are the most similar.
\citet{beckwith:08} found that simply scaling
the radial coordinate by $r_{\rm ISCO}$, i.e., using $r^\prime \equiv r/r_{\rm ISCO}$,
gives a roughly universal emissivity profile for NT disks, but not MHD disks.
To unite MHD disks around black holes of different spins, we have instead developed
a coordinate transformation based on the differential proper radial distance
$d\tilde{r}=g_{rr}^{1/2}dr$.  This coordinate transformation requires a constant of
integration, or alternatively a free parameter that allows us to
choose the radius at which $\tilde{r}=r$. We do this at
$r=\frac{3}{2}r_{\rm ISCO}$, corresponding closely to the radius of
peak emissivity. Finally, we scale by $r_{\rm ISCO}$ to make the
radial coordinate dimensionless: $r^\ast \equiv \tilde{r}/r_{\rm ISCO}$.
Note that in these coordinates, the horizon is located at negative $r^\ast$, but the
vast majority of the flux is emitted outside of $r^\ast=0$, where the Boyer-Lindquist radial coordinate is $r\approx 1.15 r_{\rm
  hor}$ for all values of $a/M$.  Details of the coordinate
transformation and corresponding emissivity law are presented in
Appendix A. 

\begin{figure}[ht]
\caption{\label{fig:dldr_rstar} Luminosity profile $dL/dr^\ast$, for a
  range of spins, all normalized to give a bolometric luminosity of
  $L=0.1L_{\rm Edd}$ for a black hole mass of $10M_\odot$. The solid
  curves correspond to the {\it emitted} 
  luminosity, i.e., the black solid curves of Fig.\
  \ref{fig:harm_nt}. The black dashed curve is the best-fit
  smoothly-broken power law (SBPL) function defined in Appendix A, and
  the dotted (dot-dashed) curve corresponds to a NT disk with $a/M=0$
  ($0.9$).}
\begin{center}
\includegraphics[width=0.8\textwidth]{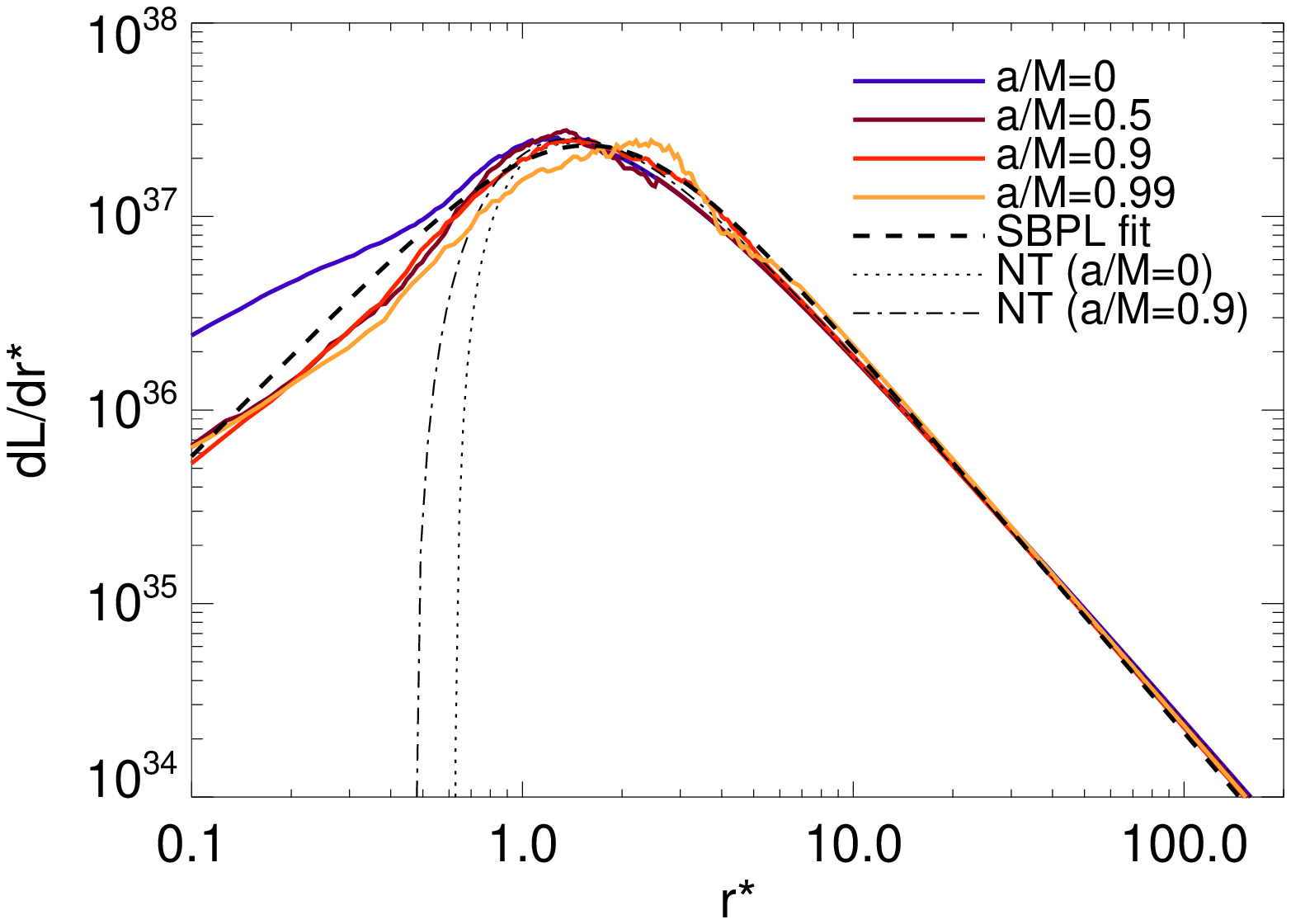}
\end{center}
\end{figure}

In Figure \ref{fig:dldr_rstar} we show the emission profiles of the
four different MHD simulations posed in terms of the dimensionless radial coordinate
$r^\ast$. As expected, they all peak around $r^\ast \approx 1.5$, and
at large $r^\ast$, they all approach the Newtonian large-radius limit
$dL/dr \propto r^{-2}$. We also show a simple analytic fit to the
emissivity profile, which takes the form of a smoothly-broken power
law with inner slope of $dL/dr^\ast\propto r^{\ast 7/4}$ (SBPL; dashed
curve), as well as the corresponding profiles for Novikov-Thorne disks
($a/M=0$ dotted curve, $a/M=0.9$ dot-dashed curve). The local flux
from the disk is related to $dL/dr^\ast$ through the relation 
\begin{equation}\label{eqn:Frstar}
F(r^\ast) = \frac{1}{4\pi r^\ast} \frac{dr^\ast}{dr}
\frac{dL}{dr^\ast}\, .
\end{equation}

Along with the emissivity profile, a simple disk model requires a
prescription for the fluid velocity. Outside the ISCO, we can use
circular, planar geodesic orbits as in \citet{novikov:73}.
Inside the ISCO, we ignore the fluid's continuing loss of energy and
angular momentum, assuming it follows
plunge trajectories with constant specific energy and angular
momentum $\varepsilon_{\rm ISCO}$ and $\ell_{\rm ISCO}$. 

\begin{figure}[ht]
\caption{\label{fig:u_r} Velocity components as a function of radius
  for spin parameter $a/M=0.5$. The solid red lines come from the time-
  and azimuth-averaged simulation data, sampled in the disk
  midplane. The diamond points correspond to the analytic model for
  circular, planar geodesic orbits outside the ISCO and plunging
  trajectories with constant energy and angular momentum.   Note the
extremely small magnitude of $u^\theta$ found in the simulation.}
\begin{center}
\includegraphics[width=0.45\textwidth]{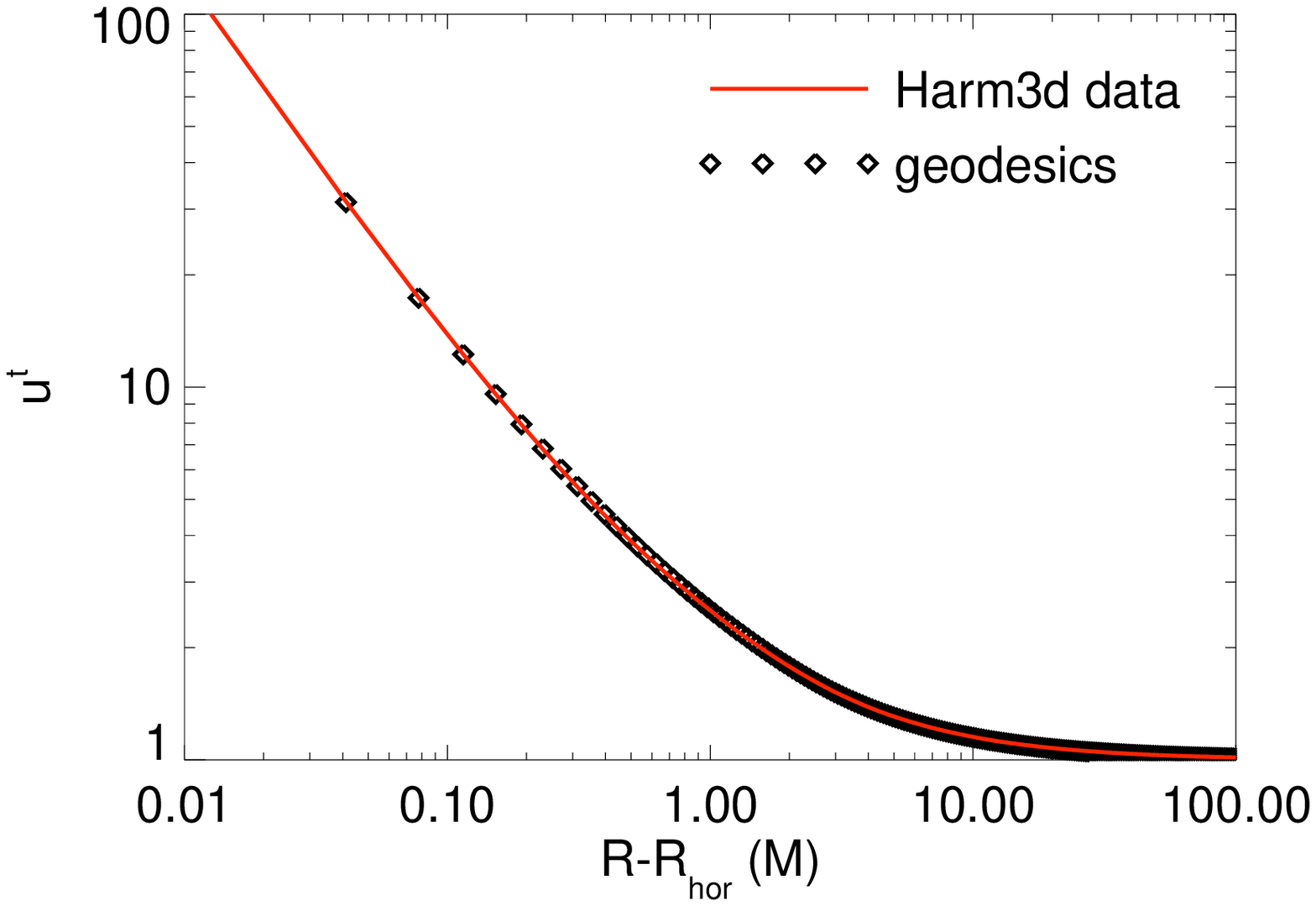}
\includegraphics[width=0.45\textwidth]{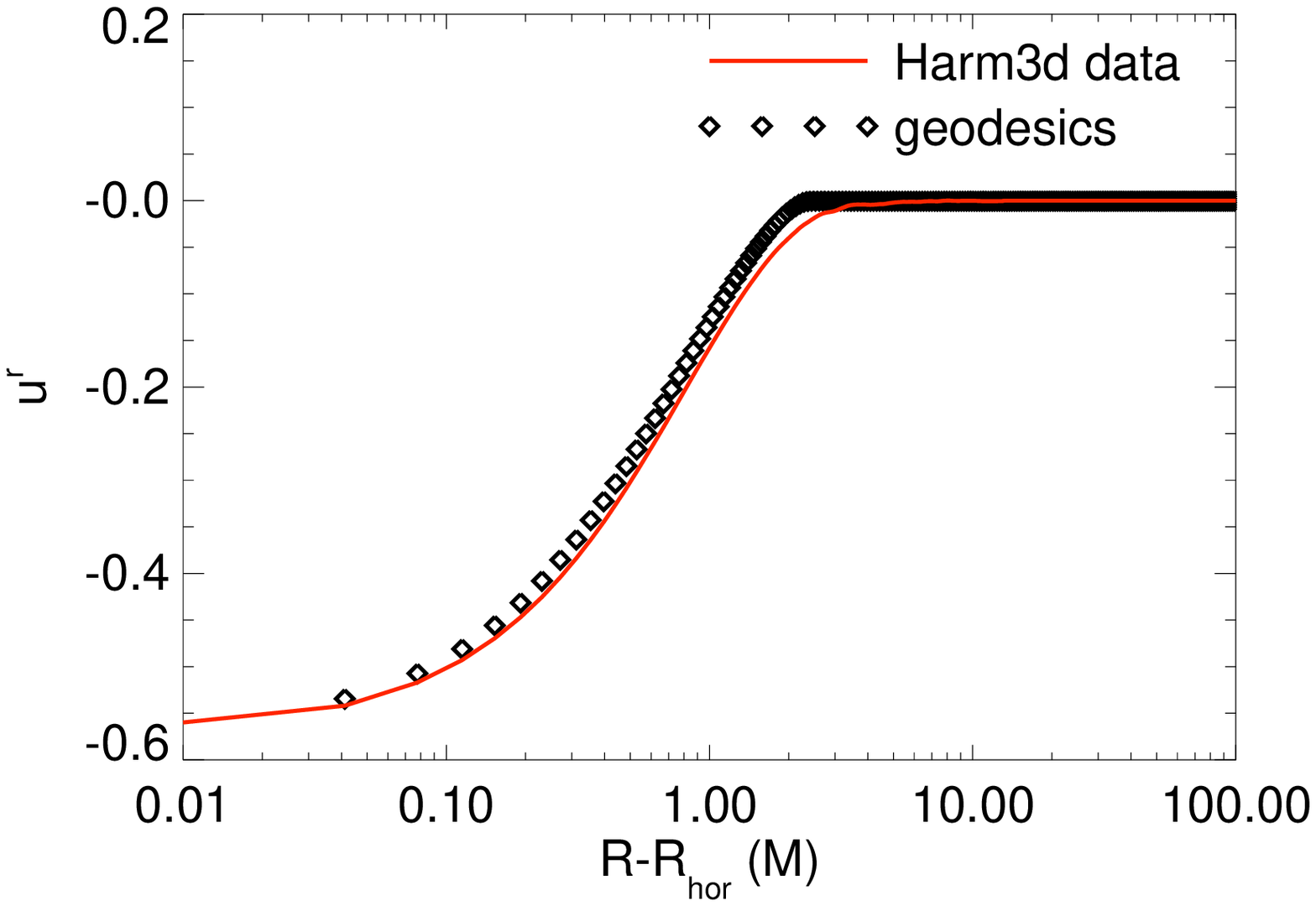}\\
\includegraphics[width=0.45\textwidth]{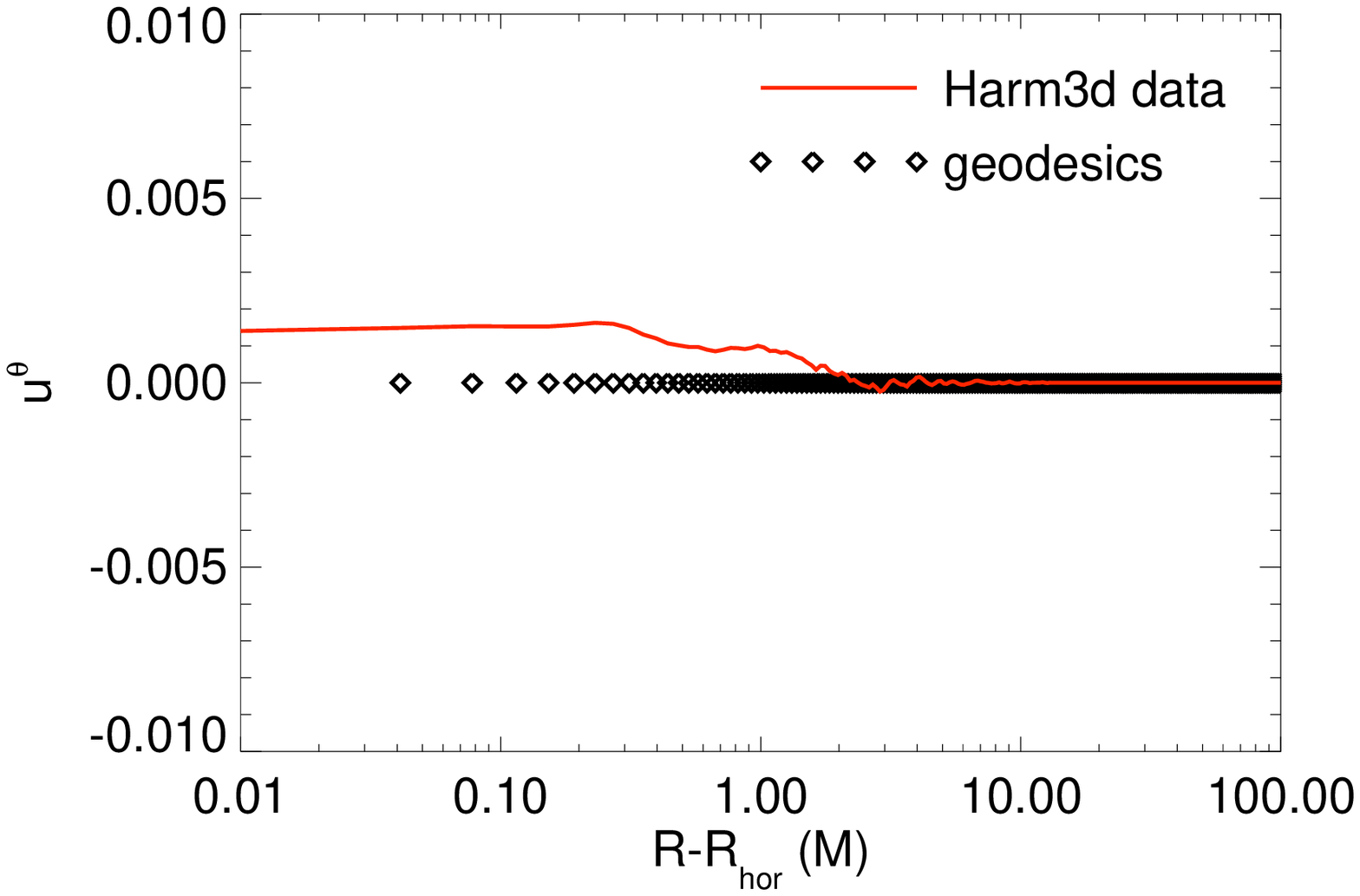}
\includegraphics[width=0.45\textwidth]{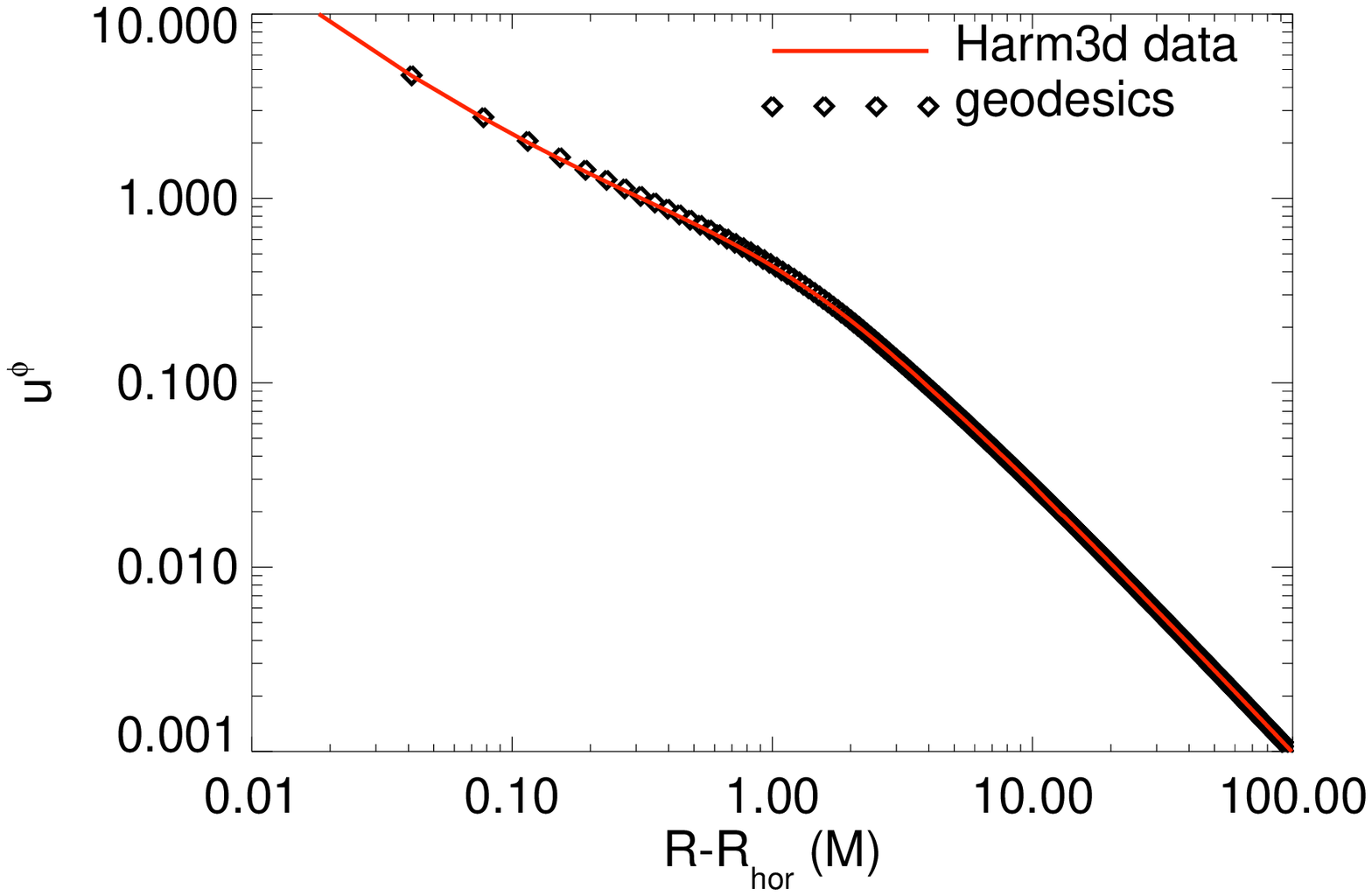}
\end{center}
\end{figure}

In Figure \ref{fig:u_r} we show how well these geodesic orbits agree
with the time- and azimuth-averaged velocities from the $a/M=0.5$ simulation
data sampled in the midplane. To emphasize the behavior in the
plunging region, we plot $u^\mu$ vs $r-r_{\rm horizon}$ (Boyer-Lindquist
coordinates). The $a/M=0.5$ case demonstrates that we
accurately capture spin effects while also including a
non-negligible plunge region; for $a/M=0.99$ almost the entire disk
(described in Boyer-Lindquist coordinates) is outside the ISCO and thus
on circular orbits. 

\begin{figure}[ht]
\caption{\label{fig:spec_fits} 
  Thermal spectra from accreting black holes
  with spins $a/M=0$ and $a/M=0.5$. In both cases,
  the black hole mass is $10M_\odot$, the accretion rate is
  $\dot{m}=0.1$ (assuming the radiative efficiency of the
  ``true'' spin), and the observer inclination is $i=60^\circ$. The
  spectra come from \harm simulation data (solid red curves) and the
  phenomenological SBPL model (dotted black curves). Novikov-Thorne
  spectra for a few representative spins are shown as thin dashed
  curves. The NT spin cases increase in luminosity with increasing spin.} 
\begin{center}
\includegraphics[width=0.45\textwidth]{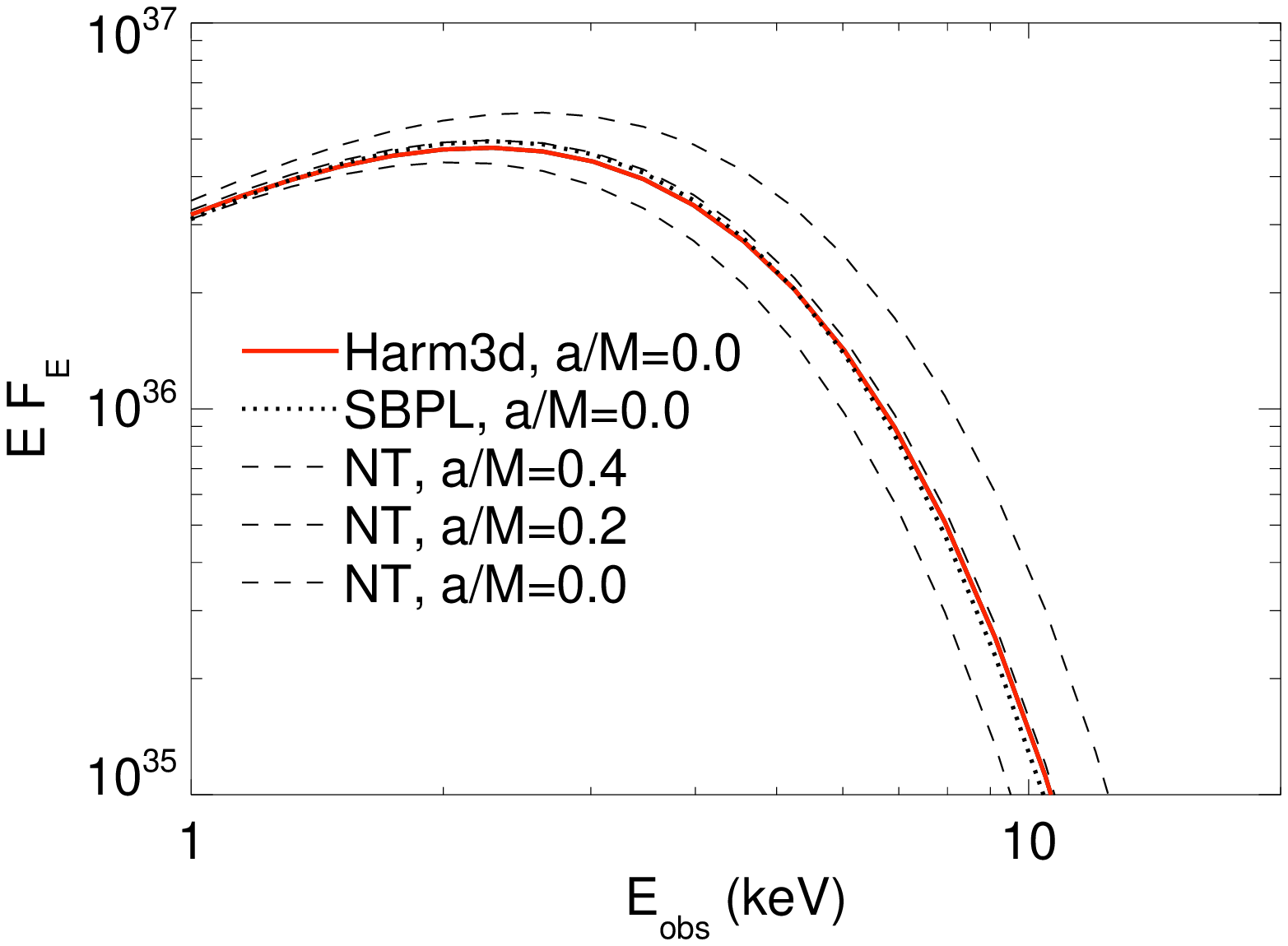}
\includegraphics[width=0.45\textwidth]{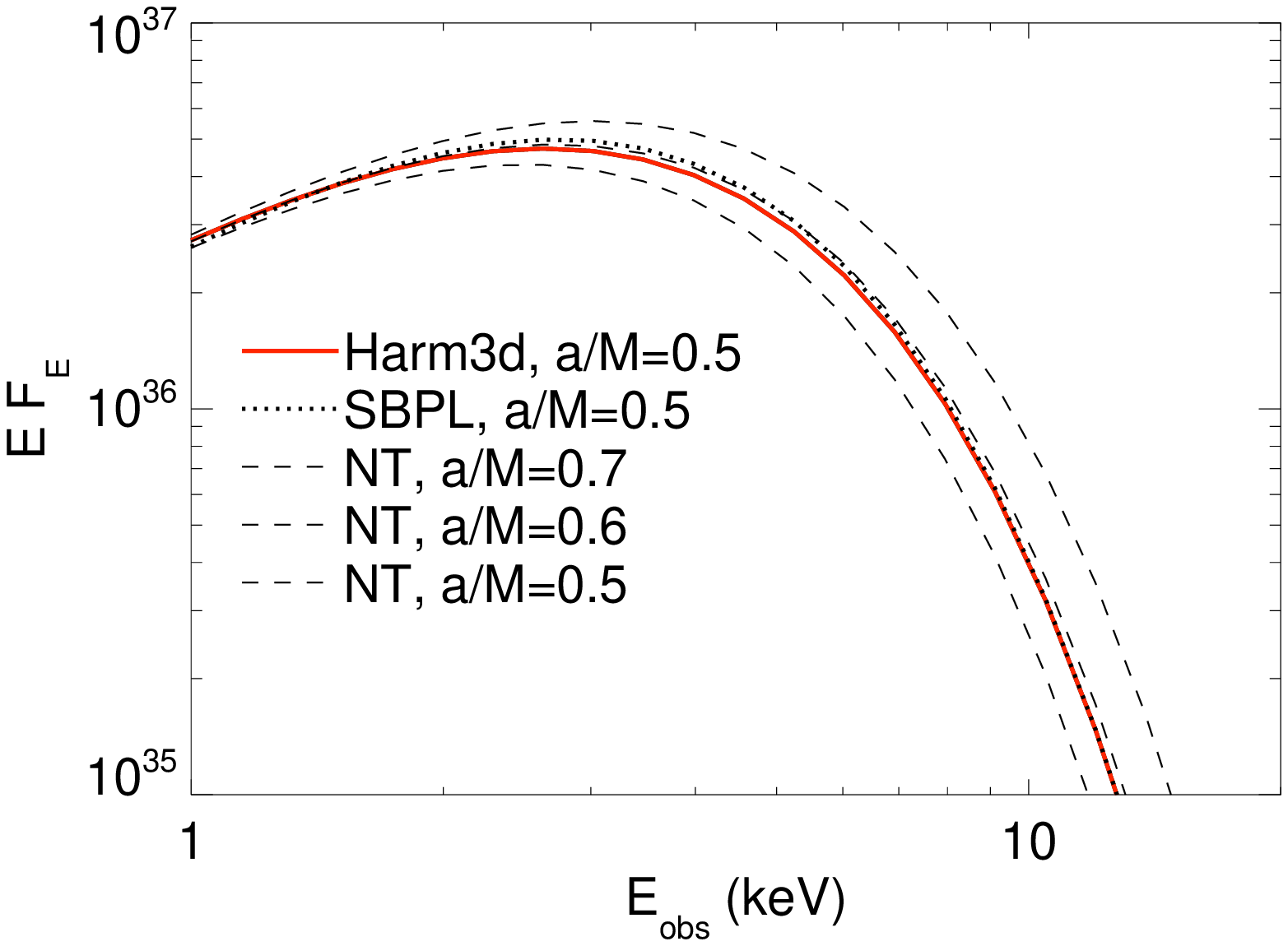}
\end{center}
\end{figure}

\section{Implications for Parameter Inferences}\label{sec:inference}

With these convenient analytic models for the thermal disks, we can
revisit the question asked in \citet{noble:11} and
\citet{kulkarni:11}, i.e., ``How best to fit data in order to infer the spin
of the black hole?''    It would be prohibitively expensive computationally
to construct a set of MHD simulations with sufficiently fine spacing in
black hole spin to support realistic data-fitting.   However, our approximate
analytic model is easily adapted to such an effort.   To validate this approach,
we first demonstrate that the spectra predicted by this model closely match
the spectra predicted from actual MHD simulations.     Given the local
emissivity and the four-velocity at each point in the disk, whether taken
from simulations or the model, we use the
relativistic radiation transport code \pan to generate spectra as seen
by a distant observer; we then compare the result.   Note that \pan includes polarization and limb-darkening
emissivity governed by electron scattering in the disk atmosphere
\citep{chandra:60}. We also include the effects of returning
radiation---photons that scatter off the opposite side of the disk
after getting deflected by the extreme gravitational lensing of the
black hole \citep{schnittman:13b}.

In Figure~\ref{fig:spec_fits} we show two examples of how well our model
reproduces actual simulation predictions for the disk's thermal spectrum, and how both
differ from the NT model. In the left panel, we show the
results from the $a/M=0$ simulation as a solid red curve, and in the
right panel, the same for $a/M=0.5$. The NT spectra for a few
comparable spins are plotted as dashed lines, and the SBPL spectra for
the ``correct'' spins are shown as dotted lines.   In both cases, the SBPL
prediction adheres very closely to the prediction directly from MHD data.
However, when the NT model uses the correct spin, it under-predicts the
luminosity at energies above the spectral peak by $\sim 10\%$. As we
will see below, when the NT model best fits, its spin is too large.

We caution the reader that these contrasts are taken assuming
all the other parameters that must be inferred from measurements
(the black hole distance $D$, mass $M$, and accretion rate $\dot{M}$)
are their correct values.   Although this cannot in general be guaranteed,
their effect on the spectrum is to alter the overall normalization and peak energy
of the spectrum, but the {\it shape} of the spectrum---especially
above the thermal peak---is a function only of the spin and
inclination. 

\subsection{$\chi^2$ fits to the spectrum:fixed inclination angle}\label{sec:spec_chisq}

To determine quantitatively how use of our new model affects parameter inference,
we adopt a simple fitting procedure in which the MHD spectrum is treated as
observational data with an error
on each energy bin proportional to the square root of the number of
photons in that bin, then carry out a least-squares fit with a suite
of NT and SBPL spectra. Rather than specify a photon count rate and
detector response function, we simply assume a total of $10^6$
photons, deposited appropriately through ten energy channels.
Initially, we hold the observer inclination fixed, but the black
hole mass, accretion rate, and distance are allowed to vary freely to
minimize $\chi^2$. The results are shown in Table
\ref{table:spin_fit_2_10} for target parameters of $M=10M_\odot$,
$\dot{m}=0.1$, and under the assumption that all the cooling power is
efficiently converted into thermal flux at the disk surface. 

\begin{table}[ht]
\caption{\label{table:spin_fit_2_10} Spin fitting, $E=2-10$ keV}
\begin{center}
\begin{tabular}{llccc}
\hline
\hline
$a/M$ & $i$ (deg) & best fit       & best fit \\
      &           & Novikov-Thorne & SBPL emission \\
\hline
0.0  & 15 & 0.20 & 0.10 \\
     & 45 & 0.20 & 0.05 \\
     & 75 & 0.30 & 0.10 \\
\hline
0.5  & 15 & 0.60 & 0.55 \\
     & 45 & 0.55 & 0.50 \\
     & 75 & 0.60 & 0.50 \\
\hline
0.9  & 15 & 0.95 & 0.87 \\
   & 45 & 0.92 & 0.85 \\
   & 75 & 0.93 & 0.87 \\
\hline
0.99 & 15 & .997 & .991 \\
  & 45 & .992 & .980 \\
  & 75 & .991 & .980 \\
\hline
\end{tabular}
\end{center}
\end{table}

These results are in agreement with the findings of \citet{noble:11},
finding a modest but consistent overestimate of the spin parameter
when fitting to a NT disk model.   \citet{kulkarni:11} also found a systematic
error due to assuming the NT model, but of slightly smaller magnitude than both
\citet{noble:11} and the new results presented here. Some possible
explanations for these differences were outlined in detail in
\citet{noble:11}. One potential distinction mentioned there was the
difference between averaging over $t$ and $\phi$ before or after
calculating the spectra. We will discuss this issue in more detail
below in Section \ref{sec:fitangle}, but for the purposes of
facilitating direct comparison with \citet{kulkarni:11}, the fits in
table \ref{table:spin_fit_2_10} were done {\it after} averaging the
simulation data over time and azimuth, then carrying out a single
ray-tracing spectral calculation. 

The fits quoted in Table \ref{table:spin_fit_2_10} were carried out
over the energy range 2--10 keV, appropriate for the large body of
archival data from black holes in the thermal state. From Figure
\ref{fig:spec_fits}, it is clear that this energy range is also ideal
for distinguishing between different spins for systems with comparable
masses and luminosities to the fiducial $M=10M_\odot$ and
$\dot{m}=0.1$. When performing the fits over a wider energy range
0.1--10 keV, the outer disk contributes greater weight, yet this part
of the spectrum is nearly independent of black hole spin, so the net
effect is almost negligible, and we find nearly identical values for
the best-fit spins. 

Not surprisingly, the SBPL model does a better job at recovering the
spin parameter for the target spectra, particularly for smaller values
of the spin, where emission from the ISCO region is relatively
more important. The small variations from
perfect fitting are likely due to the stochastic nature of the MHD
simulations as described above. There does not appear to be any clear
trend for systematic offsets with observer inclination. Taken
together, the deviations between the target spins and the best-fit
SBPL spins can be taken as a rough estimate for the systematic error
in the spin fitting method. Comparing with Table \ref{table:eta}, it
is clear that the radiative efficiency alone is not a sufficient
measure of the thermal emissivity: the MHD simulations give 
efficiencies that are very similar to NT, but the shapes of the observed
thermal spectra show deviations from the classic thin disk model. And
ultimately, it is the spectral shape that is observed, not the
radiative efficiency.

\subsection{$\chi^2$ fits to the spectrum: unknown inclination angle}\label{sec:fitangle}

\begin{figure}[ht]
\caption{\label{fig:contourJS1} Contours of $\Delta \chi^2$ for
  simulated spectral observations of an accreting black hole in the
  thermal state. The spectrum is fit in the range $2-10$ keV with
  energy resolution of $\Delta E/E \sim 0.2$. The target system
  (marked by '*') has
  spin parameter $a/M=0.5$, inclination $i=60^\circ$, mass $M=10
  M_\odot$, and luminosity $L=0.1L_{\rm Edd}$. The contours represent
  1-, 2-, and 3-$\sigma$ confidence limits.}
\begin{center}
\includegraphics[width=0.6\textwidth]{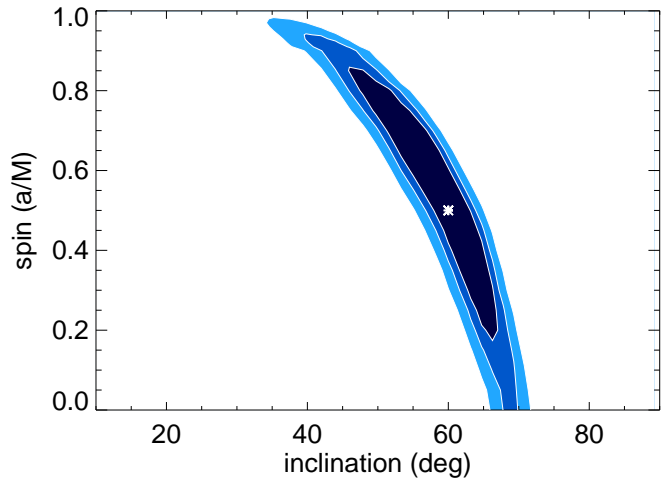}
\end{center}
\end{figure}

As shown in Figure \ref{fig:spec_fits}, the shape of the thermal
spectrum in the $2-10$ keV band is a function of the black hole
spin. For a fixed bolometric luminosity, a larger spin leads to a higher
disk temperature because the total radiative power is emitted from a
smaller region.  However, relativistic beaming and Doppler blue-shifting
of photons emitted from a highly inclined disk have a similar
hardening effect on the observed spectrum. Together, there is an
almost perfect degeneracy between the black hole spin and the observer
inclination angle. To demonstrate this, we plot in Figure
\ref{fig:contourJS1} contours of $\Delta \chi^2$ between a simulated
target spectrum generated from the MHD data with $i=60^\circ$ and
$a/M=0.5$, and a template of fitting spectra based on the SBPL
model. The simulated spectral data have Poisson error bars consistent
with an observation of roughly $10^6$ photons divided into 10 energy
bins, typical of what might be expected from a first-generation X-ray
polarimeter \citep{gems,ixpe}.

This degeneracy is a well-known problem with the continuum
fitting method, and is generally resolved by assuming a specific
inclination derived from the optical light curve of the distorted
stellar companion \citep{orosz:02,remillard:06}.
One systematic problem with this approach is that the inclination of
the inner disk may very well be misaligned with the inclination of the
binary orbit \citep{li:09}. 

Another important issue in fitting
observations is properly treating the scattering atmosphere of the
accretion disk. Because the electron scattering opacity dominates over
free-free absorption for the temperatures and densities expected in
the inner disk, we expect a Wien spectrum for the local emission
\citep{rybicki:04}. For fitting our simulated spectra to NT or SBPL
models, motivated by \citet{shimura:95} we used a simple hardening
factor of 1.8 as in \citet{noble:11}. When fitting real observational
data, \citet{mcclintock:14} find that more detailed atmospheric models are
required to achieve a good fit \citep{davis:06}.

As mentioned above in Section \ref{section:results}, there is also the
issue of t- and $\phi$-averaging the data before carrying out the
ray-tracing calculation. This
averaging has the effect of softening the observed spectrum. Consider
an annulus of the disk with a uniform temperature $T_0$. If instead,
the total thermal flux were emitted from a series of discrete hotspots
of temperature $T_h$ and covering fraction $C_h$, while maintaining
the same total flux, the new spectrum is a higher-temperature
blackbody with color-correction factor $f_c = C_h^{-1/4}$.

As a test
of this effect, we generated spectra both before and after
averaging in time and azimuth, and find that by first averaging the
simulation data and using a color-correction factor of $f_c=1.8$, we
can exactly match the spectra that result from using the full
time-varying 3D data with $f_c=1.6$ and then integrating the light
curves in time and azimuth. Of course, Nature uses the latter method,
so if the observational data best fits an azimuthally symmetric model
best with a certain hardening function, one must keep in mind that
this in fact corresponds to a somewhat softer local spectrum coming
from an inhomogeneous disk. Careful analysis of X-ray timing
properties may be one promising way of breaking this degeneracy and
lead to greater understanding of the properties of the accretion disk
atmosphere.

\subsection{$\chi^2$ fits to the X-ray polarization}\label{sec:poln}

As discussed in detail in \citet{schnittman:09}, X-ray polarization
observations of black holes in the thermal state provide a powerful
new way to probe the dynamics of the inner accretion flow, and thereby
measure the black hole spin. \citet{chandra:60} showed that the
photons emitted from a plane-parallel, scattering-dominated atmosphere
are polarized in a direction parallel to the emitting
surface. The degree of polarization ranges from 0 for face-on
observers up to $11\%$ for edge-on observers. Because of relativistic
effects such as beaming and gravitational lensing, the net
polarization of light from the inner regions of an accretion disk is
reduced in magnitude, and the angle of polarization is rotated. Because
higher-energy photons come from the inner regions, these relativistic
effects are energy dependent, and thus the polarization spectrum can
be used to probe the inner disk. 

This technique was first proposed
over thirty years ago by \citet{connors:80}, who considered only direct
radiation from the accretion disk. \citet{schnittman:09} included the
effects of returning radiation, photons emitted from the disk,
deflected by the black hole, and reflected off the far side of the
disk before reaching the distant observer.   These photons are strongly
polarized by the large-angle scattering geometry and impart a significant
polarization signature to the total radiated flux.

For the purpose of analyzing the X-ray polarization, we define the black hole
spin axis to be along the vertical direction, so
a polarization angle of $\psi=0^\circ$ is parallel to the disk. This
is the case for the low-energy photons emitted at large radii, where
relativistic effects are negligible. The high-energy photons emitted
close to the black hole are more likely to contribute to the returning
radiation and scatter into the vertical polarization mode. Since
the disks around rapidly spinning black holes extend closer to the
horizon, they should exhibit more vertical polarization, and the
transition from horizontal to vertical polarization occurs at a lower
energy. 

\begin{figure}[ht]
\caption{\label{fig:polarization} Polarization amplitude (top) and
  angle (bottom) for thermal disks as a function of photon energy for
  a range of black hole spins. In
  all cases, the black hole parameters are $M=10M_\odot$, $L=0.1L_{\rm
  Edd}$, and $i=75^\circ$.}
\begin{center}
\includegraphics[width=0.45\textwidth]{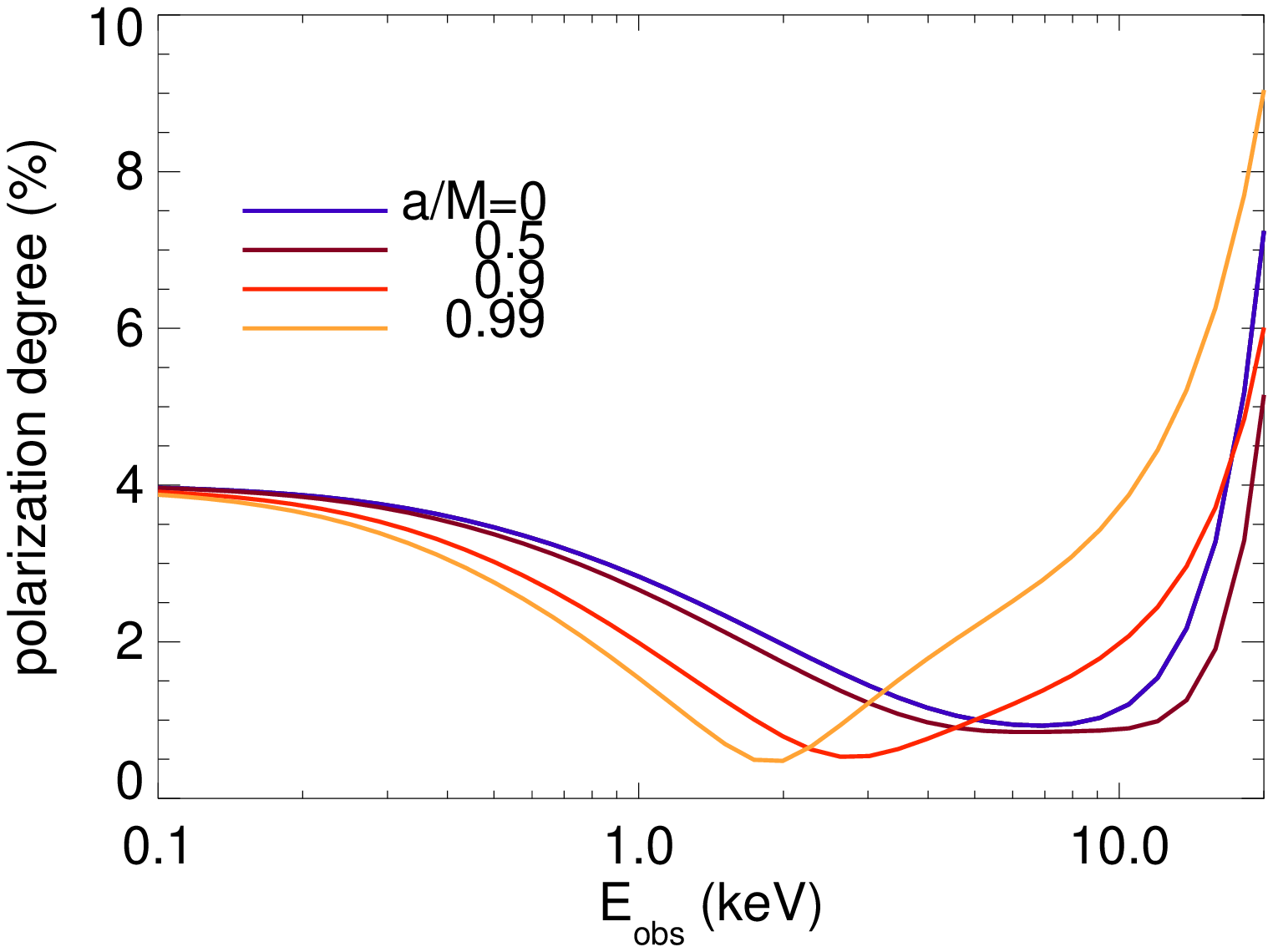}
\includegraphics[width=0.45\textwidth]{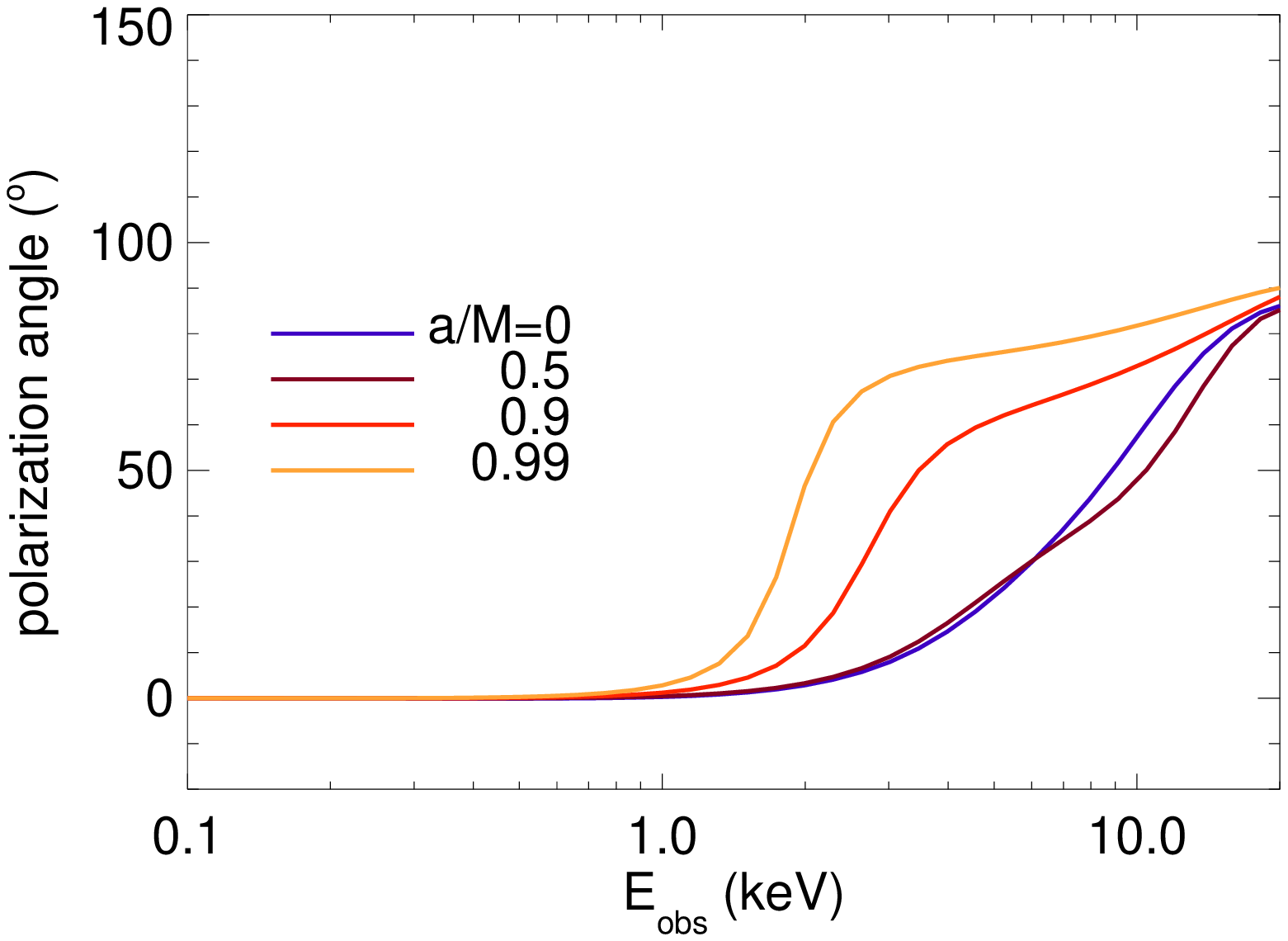}
\end{center}
\end{figure}

In Figure \ref{fig:polarization} we plot the angle and degree of
polarization from the MHD simulation data as a function of photon
energy for our fiducial parameters $M=10M_\odot$, $L=0.1L_{\rm Edd}$,
and $i=75^\circ$. The spectropolarization signal is clearly sufficient
to distinguish between different spins with even modest energy
resolution. The near overlap of the $a/M=0$ and $a/M=0.5$ curves is
due to the anomalously low (high) flux from the $a/M=0.5$ ($a/M=0$)
simulations, as described above. Real observations would cover a much
longer duration of time, smoothing out the stochastic
variability seen in the much shorter simulations.

\begin{figure}[ht]
\caption{\label{fig:contourJS2} Contours of confidence limits as in
  Figure \ref{fig:contourJS1}, but now also considering polarization
  spectra $Q_\nu$ and $U_\nu$. The simulated sensitivity corresponds
  to a minimum detectable polarization of $\sim 1\%$.}
\begin{center}
\includegraphics[width=0.6\textwidth]{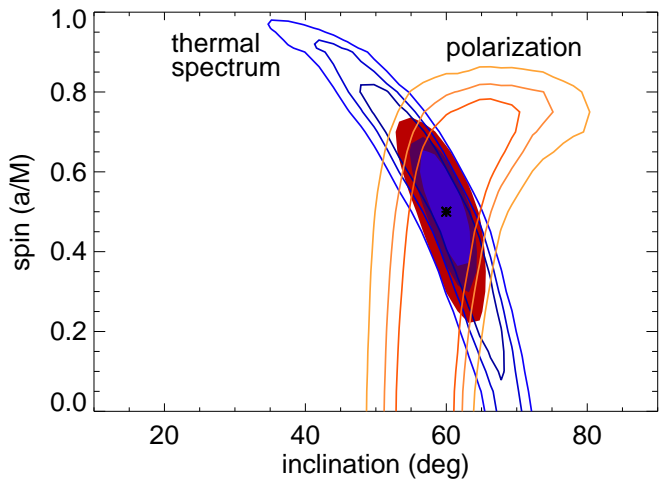}
\end{center}
\end{figure}

Polarization also provides a complementary observable that
can break the spin-inclination degeneracy inherent in the traditional spectral
continuum fitting method \citet{li:09}.   To demonstrate its power,
we perform the same sort of spectral fitting shown in Figure
\ref{fig:contourJS1}, but using the broad-band polarization information
displayed in Figure \ref{fig:polarization}.  Instead of the intensity
spectrum, we use the energy-dependent Stokes parameters $Q_\nu$ and
$U_\nu$, as in \citet{schnittman:09}. Again, we use a simulated
spectrum from a target disk with $a/M=0.5$ and $i=60^\circ$. For a
broad-band photoelectric polarimeter like PRAXyS \citep{gems}, $10^6$ photons
divided roughly equally between 10 energy bands gives a 1-$\sigma$ uncertainty
on the polarization degree of about $1\%$ per band. Fitting the
polarization information with the SBPL model between 2-10 keV, we get
the confidence contours shown in Figure \ref{fig:contourJS2}.
At 2 keV there is still significant
returning radiation, so the polarization signal is not a perfect probe
of disk inclination. A soft X-ray polarimeter sensitive over the 0.1-0.5 keV
band would significantly improve sensitivity to disk inclination. 

The slight spin-inclination degeneracy seen in Figure
\ref{fig:contourJS2} can be understood as follows. From Figure
\ref{fig:polarization} it is clear that the location of the transition
from horizontal to vertical orientation occurs at lower energy for
increasing spin. This transition is due to a small number of highly
polarized returning photons dominating over the vast majority of
weakly polarized direct photons. At higher inclinations, the direct
signal is more highly polarized, so is harder to overcome with
returning photons, and thus the transition point shifts to higher
energy where the contribution from returning radiation is higher.  Therefore,
to maintain a constant transition energy, increasing inclinations
require a higher spin. 

Comparing the contours in Figure
\ref{fig:contourJS2}, we see that the degeneracy due to continuum
fitting and that due to polarization intersect at a significant angle.   As a result,
the range of both parameters consistent with a good fit is substantially reduced.
Clearly, X-ray spectropolarimetry observations have the potential to be of
great benefit to measurements of black hole spin and inclination.

For all of the contours in Figure
\ref{fig:contourJS2}, we used the SBPL model for
the radial emissivity profile. When redoing them with a NT model as in
\citet{noble:11}, not surprisingly we get very similarly shaped
contours for $\Delta \chi^2$, but with a systematic offset towards higher
spin. However, we find that the actual minimum in $\chi^2$ is
consistently lower for the SBPL model than for NT.   For the fiducial observations
used above with $10^6$ photons over the 2-10 keV band, combined
spectral and polarization fits to the MHD data prefer the SBPL model
at roughly $90\%$ confidence. For a photon-limited (as opposed to
background-limited) detector, this significance would simply grow with
more and longer observations.   Thus, X-ray polarimetry combined with
X-ray spectroscopy can also be a significant aid in distinguishing different
emissivity profiles, thereby creating a new diagnostic of accretion dynamics
in the innermost regions of accretion disks around black holes.

\section{Summary}\label{section:summary}

We have carried out a new series of high-resolution GRMHD simulations
of thin accretion disks around spinning black holes. Using a
physically motivated cooling function to emulate radiative losses, we have
found a single broken power-law model that accurately describes the
luminosity profile of the thermal emission from these disks.   By reframing
the profile in terms of a new radial coordinate, this model can
capture the additional emissivity near the ISCO produced by MHD stresses,
but omitted by the traditional Novikov-Thorne model.   These new
profiles should more accurately fit the 2-10 keV spectra observed from
stellar-mass black holes in the thermal state, thereby reducing systematic
errors in parameter estimates of black hole spin and inclination.
Although a single power-law index for the inner radial dependence of
the surface brightness fits all our simulations quite well, it is
possible that this index might be altered if, for example, the
magnetic field topology were different. 

We have further found that the polarization induced by returning radiation
\citep{schnittman:09} offers a powerful probe for tightening
these parameter inferences.   It constrains both the spin and the inclination
in a way having little degeneracy with the spectral method.    At the same time,
it can also constrain the detailed shape of the intrinsic disk radial
luminosity profile (including the possibility of a different inner
power-law index).
Moreover, because the trajectories of the returning radiation photons pass
close to the black hole's photon orbit, this method probes as deeply into the
regime of strong-field relativistic gravity as almost any imaginable observational
method.

\appendix

\section{Emissivity Function}\label{section:appA}
Here we present a simple procedure for generating the phenomenological
emissivity profile that best matches the MHD simulations, for a range
of black hole spin parameters. 

We begin by defining a new radial coordinate that captures the effects
of increased proper distance as measured by a geodesic observer near
the black hole:
\begin{equation}\label{eqn:drtilde}
d\tilde{r} \equiv g_{rr}^{1/2} dr =
\left(1-\frac{2}{r}+\frac{a^2}{r^2}\right)^{1/2} dr\, ,
\end{equation}
with $dr$ the standard differential radial coordinate in
Boyer-Lindquist coordinates. Integrating equation (\ref{eqn:drtilde})
gives $\tilde{r}$ as a function of $r$ with the condition that they
are equal when $r=(3/2)r_{\rm ISCO})$. 

We next scale by the ISCO radius to define another coordinate
$r^\ast\equiv \tilde{r}/r_{\rm ISCO}$. With these coordinates, the
universal emissivity function can be written as a smoothly-broken
power law of the form
\begin{equation}\label{eqn:sbpl}
\frac{dL}{dr^\ast} = C r^{\ast\, \varphi} \left[
\frac{\cosh(\ln (r^\ast/R_0)/\Delta R)}{\cosh(\ln (1/R_0)/\Delta R)}
\right]^{\xi\, \Delta R}\, ,
\end{equation}
where $C$ is an overall normalization factor, $R_0$ is the location of
the power-law break, $\Delta R$ is the width of the break, and
$\alpha$ and $\beta$ are the slopes of the power law at small and
large radii respectively,
with $\varphi=(\beta+\alpha)/2$ and $\xi =
(\beta-\alpha)/2$. We fix $\beta=-2$ to give the proper behavior at
large $r$, and $C$ is determined by the total mass accretion rate. For
the remaining free parameters, we find the best simultaneous fit with
$\alpha = 1.73$, $R_0=1.68$, and $\Delta R = 0.92$. 

From equations (\ref{eqn:sbpl}) and (\ref{eqn:Frstar}) above, we can
write the local flux as 
\begin{equation}\label{eqn:FrstarA}
F(r^\ast) = \frac{1}{4\pi r^\ast} \frac{dr^\ast}{dr}
\frac{dL}{dr^\ast}\, .
\end{equation}
For purely blackbody emission, the local temperature is then given by
$T(r)=[F(r)/(2\sigma)]^{1/4}$. The normalization factor $C$ in
(\ref{eqn:sbpl}) is simply a function of the total luminosity:
$L=\int_{0}^{\infty}dr^{\ast}(dL/dr^{\ast}$ and scales linearly with
the mass accretion rate. The black hole mass $M$ comes in when converting
the dimensionless coordinate $r^\ast$ to cgs units, leading to a
factor of $M^{-2}$ in equation (\ref{eqn:FrstarA}), giving the
familiar scaling of disk temperature $T\sim M^{-1/2}\dot{M}^{1/4}$.

Outside of the ISCO, the gas moves in planar, circular orbits with
specific energy and angular momentum
\begin{subequations}
\begin{eqnarray}
\varepsilon_{\rm circ}(r) &=& \frac{r^2-2r+ar^{1/2}}{r(r^2-3r+2ar^{1/2})^{1/2}} \\
\ell_{\rm circ}(r) &=& \frac{2}{3^{1/2}}\frac{3r^{1/2}-2a}{r^{1/2}}\, .
\end{eqnarray}
\end{subequations}
In the plunging region, the gas follows geodesics with constant energy
$\varepsilon(r_{\rm ISCO})$ and angular momentum $\ell(r_{\rm
  ISCO})$. From these,
the coordinate 4-velocity is given by
\begin{subequations}
\begin{eqnarray}
u^t &=& -\varepsilon\, g^{tt} + \ell\, g^{t\phi} \\
u^r &=& u_r g^{rr} \\
u^\theta &=& 0 \\
u^\phi &=& -\varepsilon\, g^{t\phi} + \ell\, g^{\phi\phi}
\end{eqnarray}
\end{subequations}
where we can solve for the radial momentum $u_r$ from the mass-shell
normalization condition:
\begin{equation}
u_r = -\sqrt{
\frac{1+\varepsilon^2 g^{tt}-2\varepsilon \ell g^{t\phi}}{g^{rr}}}\, .
\end{equation}

This simple procedure works well to construct physically-motivated
models for thin accretion disks for all spins up to $a/M\approx
0.997$. Above that limit, the coordinate $r^{\ast}$ takes on negative
values outside of the ISCO. For extreme spins above this value, we
recommend following the traditional Novikov-Thorne emissivity models,
which in any case have almost no remaining plunging region between the
ISCO at $r=1.28M$ and the horizon at $r=1.08M$. 

\acknowledgements{
We would like to thank T.\ Kallman for helpful discussions.
This work was partially supported by NASA grants NNX14AB43G and
ATP12-0139, and NSF grant AST-0908336.}

\end{document}